\documentclass[12pt]{article}
\usepackage{latexsym}
\usepackage{amsmath}
\usepackage{amssymb}
\catcode`\@=11
\newcount\hour
\newcount\minute
\newtoks\amorpm \hour=\time\divide\hour by 60\minute
=\time{\multiply\hour by 60 \global\advance\minute by-\hour}
\edef\standardtime{{\ifnum\hour<12 \global\amorpm={am}%
        \else\global\amorpm={pm}\advance\hour by-12 \fi
        \ifnum\hour=0 \hour=12 \fi
        \number\hour:\ifnum\minute<10
        0\fi\number\minute\the\amorpm}}
\edef\militarytime{\number\hour:\ifnum\minute<10
0\fi\number\minute}
\def\draftlabel#1{{\@bsphack\if@filesw {\let\thepage\relax
   \xdef\@gtempa{\write\@auxout{\string
      \newlabel{#1}{{\@currentlabel}{\thepage}}}}}\@gtempa
   \if@nobreak \ifvmode\nobreak\fi\fi\fi\@esphack}
        \gdef\@eqnlabel{#1}}
\def\@eqnlabel{}
\def\@vacuum{}
\def\marginnote#1{}
\def\draftmarginnote#1{\marginpar{\raggedright\scriptsize\tt#1}}
\def\draft{
        \pagestyle{plain}
        \overfullrule=2pt
        \oddsidemargin -.1truein
        \def\@oddhead{\sl \phantom{\today\quad\militarytime} \hfil
        \smash{\Large\sl DRAFT} \hfil \today\quad\militarytime}
        \let\@evenhead\@oddhead
        \let\label=\draftlabel
        \let\marginnote=\draftmarginnote
        \def\ps@empty{\let\@mkboth\@gobbletwo
        \def\@oddfoot{\hfil \smash{\Large\sl DRAFT} \hfil}
        \let\@evenfoot\@oddhead}
        \def\@eqnnum{(\theequation)\rlap{\kern\marginparsep\tt\@eqnlabel}%
        \global\let\@eqnlabel\@vacuum}  }

\renewcommand{\theequation}{\thesection.\arabic{equation}}
\renewcommand{\thefootnote}{\fnsymbol{footnote}}
\newcommand{\newsection}{    
\setcounter{equation}{0}\section}
\def\appendix#1{\addtocounter{section}{1}\setcounter{equation}{0}
\renewcommand{\thesection}{\Alph{section}}
\section*{Appendix \thesection\protect\indent \parbox[t]{11.15cm}{#1}}
\addcontentsline{toc}{section}{Appendix \thesection\ \ \ #1}}

\def \bi{\bibitem}
\def \la {\label}

\def \b {\beta}

\def \d {\partial}

\def\be{\begin{equation}}
\def\ee{\end{equation}}

\catcode`\@=11

\def\bea{\begin{eqnarray}}
\def\eea{\end{eqnarray}}
\def\beann{\begin{eqnarray*}}
\def\eeann{\end{eqnarray*}}
\def\beq{\begin{equation}}
\def\eeq{\end{equation}}
\def\ba{\begin{array}}
\def\ea{\end{array}}
\def\ben{\begin{enumerate}}
\def\een{\end{enumerate}}
 \def \l {\lambda}

 \def \la {\label}
 \def\be{\begin{equation}}
\def\ee{\end{equation}}

\def \la {\label}

\font\mybb=msbm10 at 11pt

\def\bb#1{\hbox{\mybb#1}}

\def\bR {\bb{R}}

\def\bC {\bb{C}}

\def\e  {\epsilon}

\def \ee {\epsilon}

\def \g {\gamma}
\def \bi{\bibitem}
\def\a{\alpha }

\def \d {\delta}

\def \l {\lambda}

\def \g {\gamma}

\def \b {\beta}

\def\be{\begin{equation}}
\def\ee{\end{equation}}

\def \bi {\bibitem}
\def \la{\label}

\def \cF {{\cal{F}}}
\def \cL {{\cal{L}}}

\begin{document}
\date{November 2002}
\begin{titlepage}
\begin{center}

\vspace{2.0cm} {\Large \bf   IIB backgrounds with five-form flux}
\\[.2cm]

\vspace{1.5cm}
 {\large  U. Gran$^1$, J. Gutowski$^2$ and G. Papadopoulos$^3$}

\vspace{0.5cm}

${}^1$ Fundamental Physics\\
Chalmers University of Technology\\
SE-412 96 G\"oteborg, Sweden\\

\vspace{0.5cm}
${}^2$ DAMTP, Centre for Mathematical Sciences\\
University of Cambridge\\
Wilberforce Road, Cambridge, CB3 0WA, UK

 \vspace{0.5cm}
${}^3$ Department of Mathematics\\
King's College London\\
Strand\\
London WC2R 2LS, UK\\
\end{center}

\vskip 1.5 cm
\begin{abstract}

We investigate all $N=2$ supersymmetric IIB supergravity
backgrounds with  non-vanishing five-form flux.
The Killing spinors have stability subgroups $Spin(7)\ltimes\bR^8$,
$SU(4)\ltimes\bR^8$ and $G_2$. In the $SU(4)\ltimes\bR^8$ case, two different
types of geometry arise depending
on whether  the Killing spinors are generic or pure.
 In both cases, the backgrounds admit a  null Killing vector field which
 leaves invariant the $SU(4)\ltimes \bR^8$ structure,
  and an almost complex structure in the directions
 transverse to the lightcone. In the generic case, the twist of the vector
 field is trivial but the almost complex structure is non-integrable, while
 in the pure case the twist is non-trivial but the almost complex structure
 is integrable and associated with a relatively balanced Hermitian structure.
The $G_2$ backgrounds admit a time-like Killing vector field and two
spacelike closed one-forms, and the seven directions transverse
to these admit a co-symplectic $G_2$ structure.
 The  $Spin(7)\ltimes\bR^8$ backgrounds are pp-waves propagating
 in an eight-dimensional manifold with holonomy $Spin(7)$.
In addition we show that all the supersymmetric solutions of simple
five-dimensional supergravity with a time-like Killing vector field,
which include the $AdS_5$ black holes, lift to $SU(4)\ltimes\bR^8$ pure
Killing spinor IIB backgrounds. We also show that the LLM solution is
associated with a co-symplectic co-homogeneity one $G_2$ manifold
which has principal orbit $S^3\times S^3$.

\end{abstract}
\end{titlepage}
\newpage
\setcounter{page}{1}
\renewcommand{\thefootnote}{\arabic{footnote}}
\setcounter{footnote}{0}

\setcounter{section}{0}
\setcounter{subsection}{0}
\newsection{Introduction}

Supersymmetric IIB backgrounds with ``active''  five-form fluxes and with vanishing one- and three-form
field strengths have been extensively investigated
in the context of string theory, branes and black holes. Examples of such backgrounds are the Freund-Rubin space
$AdS_5\times S^5$ \cite{schwarz}, the D3-brane \cite{d3brane} and the maximally supersymmetric plane  wave \cite{bfhp} solutions  which
have been instrumental in the formulation and understanding of the AdS/CFT correspondence \cite{maldacena}.
More recently, in the same context many
new solutions of increasing complexity have been
found  preserving some supersymmetry. These include the bubbling solutions of \cite{llmsol} and
the lift of $AdS_5$ black holes \cite{hjg} to IIB supergravity \cite{jgjg, ahn}.

Motivated by the above  widespread applications, we  present a systematic
investigation of all supersymmetric IIB supergravity backgrounds with active  five-form flux $F$. This is based on our
  solution \cite{ugjggpa, ugjggpb} of the Killing spinor
equations of  IIB supergravity \cite{sw, schwarz, howe} for one Killing spinor,  using the spinorial geometry
technique of \cite{jguggp}.
Although, the supersymmetric backgrounds with $F$ flux
are special cases of the $N=1$ IIB backgrounds, there are some differences.
Unlike generic IIB supersymmetric backgrounds, backgrounds with (only) $F$ flux always preserve
an even number of supersymmetries.
So the backgrounds we shall investigate will preserve at least two supersymmetries, $N\geq2$.
In addition, the vanishing of one- and three-form fluxes imposes additional
conditions on the geometry. It turns out that the geometry of the supersymmetric backgrounds with $F$ flux
 is rather restricted and
the five-form field strength $F$  takes a simple form.

In analogy with the generic $N=1$ IIB backgrounds, the $N=2$ supersymmetric IIB backgrounds with $F$ flux can be separated
into three classes distinguished by the stability subgroups of the Killing spinors in $Spin(9,1)$. These
are  $Spin(7)\ltimes\bR^8$, $SU(4)\ltimes\bR^8$ and $G_2$. We find that
the $Spin(7)\ltimes\bR^8$ backgrounds are pp-waves with  rotation and  null $F$ flux propagating on a holonomy $Spin(7)$
manifold whose metric depends on a wave profile coordinate.

The geometry of the $SU(4)\ltimes\bR^8$ backgrounds is more subtle. These are further  divided into two subclasses,
the generic and the pure spinor backgrounds, which have distinct geometries. The Killing spinors
of the former backgrounds  do  not obey additional conditions
apart from those imposed by $SU(4)\ltimes\bR^8$ invariance up to a possible conjugation
with a $Spin(9,1)$ transformation. The Killing spinors of the latter are pure $SU(4)\ltimes\bR^8$-invariant  spinors.
One consequence of this is that the stability subgroups of the spinors in $Spin(9,1)$
are not sufficient to characterize uniquely the geometry of the supersymmetric backgrounds. The geometry
rather depends on the embedding of the Killing spinor bundle into the spinor bundle of IIB supergravity
up to a $Spin(9,1)$ rotation as has been explained in \cite{iibsyst}.
 In both cases, the spacetime admits a
null Killing vector field $X$ with non-vanishing twist or rotation, and an almost complex structure with compatible
(4,0)-form leading to an $SU(4)$ structure in the directions transverse to the lightcone.
In the generic case, the twist takes values in $\bR^8$ and so it is trivial, the almost complex structure is not integrable, the
$W_1$, $W_4$ and $W_5$ classes associated with the $SU(4)$ structure are determined in terms of functions of the spacetime, and $W_2$ is related to $W_3$.
In the pure spinor case, the twist takes values in
 $\mathfrak{su}(4)\oplus_s\bR^8$ and so it is not trivial, the almost complex structure is integrable, i.e.~$W_1=W_2=0$, $W_4=W_5$ is given
 in terms of the twist
 of $X$, and $W_3$ is not restricted by the Killing spinor equations. We refer to these conditions on the $W$ classes\footnote{A balanced Hermitian
 structure is one for which the Lee form $W_4$ of the Hermitian form $\omega$ vanishes. In the present context,
 it is the difference of the Lee forms constructed from  the fundamental $SU(4)$ forms $\omega$ and  ${\rm Re}\chi$ that vanishes.} as a {\it relatively balanced
 $SU(4)$ Hermitian structure}.
If one imposes the additional condition that the twist of $X$
 is trivial, then the geometric conditions can be re-expressed as $d (e^H \omega^3)=d(e^H \chi)=0$, where $\omega$ and $\chi$ are
 the fundamental $SU(4)$ forms and $H$ is a spacetime function.
 In both cases, most of the components of the five form field strength are determined in terms of the geometry. The field equations
 that remain to be imposed to find solutions are the $E_{--}$ component of the Einstein equations, and the Bianchi identity of $F$.
Examples of IIB solutions that admit a pure Killing spinor are the D3-brane and
its intersections as well as the solutions which are obtained
from uplifting all $1/4$-supersymmetric solutions of
minimal five-dimensional supergravity
for which the Killing spinor generates a timelike Killing vector.
We show how the constraints on the five-dimensional
solutions obtained in \cite{gutgaunfive}
are sufficient to ensure that the pure spinor constraints in IIB
supergravity are satisfied when the solution is uplifted using
the Ansatz given in \cite{chamemp}. The null Killing vector field $X$ has trivial twist for the
 D3-branes and their intersections,  while  $X$ has non-trivial twist for the uplifts of
five-dimensional solutions.

The tangent space of IIB backgrounds  with $G_2$-invariant spinors
is the orthogonal sum of the trivial bundle of rank three and
a vector bundle of rank seven corresponding to the ``transverse directions''.
One of the directions along the trivial bundle is a  time-like
Killing vector field which also leaves invariant the $G_2$
structure, and the duals of the other two directions are
associated with closed  spacelike
 one-forms. The seven  transverse directions admit   a
 {\it co-symplectic or co-calibrated $G_2$ structure}, i.e.~$X_2=X_4=0$ in terms of $G_2$ classes.
 Moreover $X_1$, which is in the trivial representation of   $G_2$,   is expressed in terms of the covariant derivatives
 of the closed one-forms. The class $X_3$ is not restricted by the Killing spinor equations. All the components
 of $F$ are expressed in terms of the geometry. In addition the Bianchi identity of $F$ implies all the field equations.
 One of the consequences of the above geometric properties is that not all Killing spinor vector bilinears are Killing.
 This has also been the case for other $N=1$ IIB backgrounds.

 We  use the relation between supersymmetry  and geometry that we have described   to propose a
 constructive method of finding IIB solutions utilizing families of $G_2$ co-symplectic manifolds. As an example
 we explore such a construction based on the classification of co-symplectic $G_2$ manifolds of co-homogeneity one
 \cite{swann}.
We also uncover the co-symplectic geometry of the bubbling $AdS$ solutions of \cite{llmsol}.
In particular, we show
that these are  associated with a special family   of co-homogeneity one co-symplectic $G_2$ manifolds  that preserves an $SO(4)\times SO(4)$
symmetry whose principal orbit
is $S^3\times S^3$.

This paper is organized as follows: In section two, we describe the general geometric properties
of the supergravity backgrounds and define the ``transverse spaces'' of the spacetimes with Killing spinors
which have compact or non-compact stability subgroups in $Spin(9,1)$. In section three, we solve the Killing spinor equations
of generic $SU(4)\ltimes \bR^8$ backgrounds and describe their geometry. In section four, we solve the
Killing spinor equations of backgrounds that admit a pure Killing spinor,  and show that D3-branes and their
intersections as well as the uplifts of supersymmetric solutions of minimal five-dimensional
gauged supergravity admitting a timelike Killing vector field
are examples of such backgrounds. These include the $AdS_5$ black holes. In section five, we show that the $Spin(7)\ltimes \bR^8$
backgrounds are pp-waves. In section six, we solve the Killing spinor equations of $G_2$ backgrounds and relate their geometry to co-symplectic
$G_2$ manifolds. In section seven, we describe how families of co-calibrated $G_2$ manifolds can be used
to construct solutions of IIB supergravity with emphasis on those of co-homogeneity one. We find
that the bubbling $AdS$ solutions are such an  example. In section eight, we give our conclusions. In appendix A, we present the linear systems
associated with the $N=2$ IIB  supersymmetric backgrounds. In appendix B and C, we summarize some results on null and $G_2$
structures in ten-dimensions.

\newsection{Geometry and supersymmetry}

The Killing spinors of IIB supergravity are  complex positive chirality Weyl spinors, $S^+_{\bC}$.
 In the absence of one-form  $P$ and three-form $G$ field strengths, $P=G=0$, the Killing spinor equations
are linear over the complex numbers. This means that if $\epsilon$ is a Killing spinor,
then $i\epsilon$ is also Killing. Therefore IIB   backgrounds with non-vanishing five-form flux preserve an even number
of supersymmetries. This in particular
implies that if one sets $P=G=0$ in the $N=1$ backgrounds of \cite{ugjggpa} and  \cite{ugjggpb},
one can obtain the conditions for the most general $N=2$ backgrounds with five-form $F$ fluxes.
As a consequence, there are three classes of $N=2$ supersymmetric backgrounds with $F$ fluxes distinguished by the stability
subgroups $Spin(7)\ltimes\bR^8$, $SU(4)\ltimes\bR^8$ and $G_2$ of the Killing spinors in $Spin(9,1)$.

The maximal number of $SU(4)\ltimes\bR^8$- and $G_2$-invariant  spinors in $S^+_{\bC}$ is four. If the
structure group of spacetime is one of these groups, then it admits a rank four subbundle ${\cal I}$ of the spin bundle
$S^+_{\bC}$ spanned by the invariant spinors. Since we are investigating backgrounds that admit
two $SU(4)\ltimes\bR^8$- or $G_2$-invariant Killing spinors, the Killing spinor
bundle ${\cal K}$ is a subbundle of ${\cal I}$. Choosing a basis of spinors in ${\cal I}$, one can write the embedding
of ${\cal K}$ in ${\cal I}$. As we shall demonstrate confirming the analysis in \cite{ugjggpb}, the conditions
on the geometry of spacetime imposed by supersymmetry depend on the embedding of ${\cal K}$ in ${\cal I}$ up to
a $Spin(9,1)$ automorphism of  $S^+_{\bC}$. Before we proceed, we shall explain how the geometries of the different embeddings of ${\cal K}$ in ${\cal I}$
can be identified. Although the spinors in ${\cal I}$ are not Killing, nevertheless they are well-defined  on the spacetime because of the
reduction of its structure group. Consequently, one can use
a basis in ${\cal I}$ to construct the  form spinor bi-linears that describe the $SU(4)\ltimes \bR^8$
structure of the spacetime. As we shall see the restrictions on the geometry of the spacetime
that arise from the Killing spinor equations can be written as conditions on the covariant derivatives  of these bi-linears. Moreover,
the Killing spinor equations imply that (some of) the components of the five-form flux are also written in terms of these
bi-linears and their covariant derivatives.

 Killing spinors with isotropy group $K\ltimes\bR^8$, $K=Spin(7), SU(4)$, are associated with a null Killing vector field $X$.
 In the complement of the zero locus of $X$, the cotangent bundle $T^*M$  of the spacetime $M$, ${\rm dim}\,M=10$, admits a real trivial
rank one  null subbundle $I$ spanned by the associated one-form $\kappa$ to $X$, and a subbundle $P=\{\a\in \Gamma(T^*M)\,\vert\, i_X\a=0\}$ of rank 9. Moreover,
one has that
\bea
0\rightarrow I\rightarrow P\rightarrow {\cal T}^\star\rightarrow 0~
\eea
where ${\cal T}^\star$ is of rank 8 and its dual ${\cal T}$ is identified as the bundle of the ``transverse directions''
to the lightcone. Observe that $P$ is not canonically decomposed in ${\cal T}^\star$ and $I$. It is also possible
to define the bundle of higher-degree ``transverse'' forms of the spacetime. We shall not explain this
construction further here because it can be found in the appendices of \cite{typeI}.

A basis of $SU(4)\ltimes\bR^8$-invariant Killing spinors is $\{1, e_{1234}\}$
and it can be shown, up to a local $Spin(9,1)$ transformation\footnote{As we shall see there is some residual  symmetry left which we  use
to simplify the spinors further.}, that the
 $SU(4)\ltimes\bR^8$-invariant Killing spinors $(\epsilon, i\epsilon)$
are given by
\bea
\epsilon=(f-g_2+ig_1) 1+(f+g_2+ig_1) e_{1234}~,~~~f, g_2\not=0~,
\la{kssu4}
\eea
where $f, g_1, g_2$ are real  spacetime functions which describe the embedding of ${\cal K}$ in ${\cal I}$.
Our spinor conventions can be found in \cite{ugjggpa, ugjggpb}. If $g_2=0$, then the spinor is $Spin(7)\ltimes\bR^8$
invariant. We shall show that there are two classes of $N=2$ backgrounds with $SU(4)\ltimes\bR^8$-invariant
Killing spinors. One such class are the {\it generic} backgrounds for which there is no restriction
on the spacetime functions $f, g_1$ and $g_2$, i.e.~the embedding of ${\cal K}$ in ${\cal I}$ is generic.
However, there is another class of supersymmetric backgrounds with different geometry for  which $g_1=0$ and $f=\pm g_2$.
The Killing spinors are pure, i.e.~they are annihilated by a maximally isotropic subspace. Since the geometry of these two classes is different,
 the isotropy group  of the spinors in $Spin(9,1)$ and the number of supersymmetries are not sufficient
to characterize the supersymmetric backgrounds.
The $N=2$ backgrounds with $Spin(7)\ltimes\bR^8$-invariant Killing spinors can be thought of as
a special case of $N=2$ backgrounds with $SU(4)\ltimes\bR^8$-invariant Killing spinors.
As we have mentioned, these arise by setting  $g_2=0$ in (\ref{kssu4}).

It is clear from the results of \cite{ugjggpb} that if the isotropy group of the Killing spinors is $G_2$, the tangent bundle
of the spacetime decomposes as
\bea
T^*M=I^3\oplus {\cal T}^*~,
\eea
where $I^3$ is the trivial vector bundle  of rank three. Moreover, one direction in $I^3$ is spanned by a time-like
Killing vector field.  As in the previous case,  ${\cal T}$, which has rank 7, is the
``transverse'' bundle or the ``transverse directions'' of the spacetime. Unlike the previous cases, the decomposition of
$T^*M$ is the orthogonal decomposition with respect to the spacetime metric. However, it is not always the case that there is a submanifold
$B$
in $M$ such that the restriction of ${\cal T}$  on $B$ is its tangent bundle.

One can take the rank 4 bundle ${\cal I}$ of $G_2$ invariant spinors to be spanned by $(1+e_{1234}, e_{51}+e_{5234})$. Moreover,
up to a $Spin(9,1)$ gauge transformation, the  $G_2$-invariant Killing spinors of $N=2$ backgrounds can be written as
\bea
\epsilon=f(1+e_{1234})+i g(e_{51}+e_{5234})~,~~~f,g\not=0
\eea
where $f,g$ are real  spacetime functions. In this case, all the embeddings of ${\cal K}$ in ${\cal I}$
give the same geometry.

\newsection{Generic $N=2$ $SU(4)\ltimes\bR^8$ backgrounds}
\la{sgsu4}

\subsection{Conditions on the geometry}

 The linear system associated with the Killing spinor equations in this case
has been presented and solved in appendix \ref{lsgsu4}. The solution has been found  using the property that for these
backgrounds the functions
$f,g_1, g_2$ which determine the Killing spinors are generic. Here, we shall investigate the consequences
that the supersymmetry conditions have on the geometry.
To analyze the  geometry and fluxes, we introduce the pseudo-hermitian frame $(e^+, e^-, e^\a, e^{\bar \a})$, $\a=1,2,3,4$,  adapted
to the description of spinors in terms of forms and write the metric and fluxes as
\bea
&&ds^2=2 e^- e^++ \delta_{ij} e^i e^j= 2(e^- e^++ \delta_{\a\bar\b} e^\a e^{\bar\b})~,~~~
\cr
&&F= e^+\wedge \Phi+ e^-\wedge \Psi+ e^+\wedge e^-\wedge {\cal X}+{}^*[e^+\wedge e^-\wedge {\cal X}]~,
\la{mfsu4}
\eea
where $\Phi$ is a anti-self dual, and $\Psi$ is a self-dual, four-form in the eight directions
transverse to
the light-cone directions, and ${\cal X}$ is a three-form, $i,j=1,2,3,4,6,7,8,9$. The last term in the expression for $F$
is required by the self-duality\footnote{Our form conventions are ${}^*G_{A_1\dots A_{10-\ell}}={1\over \ell !}
G_{B_1\dots B_\ell}\, \e^{B_1\dots B_\ell}{}_{A_1\dots A_{10-\ell}}$, the spacetime volume form is
 $d{\rm vol}_M=e^0\wedge\dots \wedge e^9$, and the orientation of the transverse directions is given by
 $d{\rm vol}=e^1\wedge \dots \wedge e^4\wedge e^6\wedge\dots\wedge e^9$.} of $F$, ${}^*F=F$,  and it is completely determined by the spacetime metric
and ${\cal X}$.

Choosing a basis $\{1, e_{1234}\}$ in the space of $SU(4)\ltimes\bR^8$-invariant spinors ${\cal I}$ , one can show that the spacetime admits
the form spinor bi-linears
\bea
e^-~,~~~e^-\wedge \omega~,~~~e^-\wedge \chi~,
\la{bisu4}
\eea
where
\bea
&&\omega=-e^1\wedge e^6-e^2\wedge e^7-e^3\wedge e^8 -e^4\wedge e^9=-i \delta_{\a\bar\b} e^\a\wedge e^{\bar \b}~,~~~
\cr
&&\chi=(e^1+ie^6)\wedge (e^2+ie^7)\wedge (e^3+i e^8)\wedge (e^4+i e^9)~,
\la{ffsu4}
\eea
 are the fundamental hermitian and (4,0) $SU(4)$ forms, respectively.

To continue, it is convenient to carry out the description of the geometry in the gauge $f^2+g_1^2+g_2^2=1$. Moreover, it is convenient
to separate the conditions that arise from supersymmetry into those that involve the light-cone directions and those that involve the
 transverse directions only. To describe the former, let
$(e_+, e_-, e_i)$ be the co-frame of $(e^+, e^-, e^i)$, $e^A(e_B)=\delta^A{}_B$.   Then
the conditions in  \ref{lsgsu4} can be rewritten as
 \bea
{\cal L}_X g=0~,~~~ {\cal L}_X (e^-\wedge \omega)={\cal L}_X(e^-\wedge \chi)=0~,~~~de^-={1\over2} e^-\wedge dH~,
\la{ggsu1}
 \eea
where $X=e_+$ and $H=\log(1-4f^2 g_2^2)$. Therefore $X$ is Killing and leaves the $SU(4)\ltimes \bR^8$ structure invariant. The last condition
in (\ref{ggsu1}) follows from the torsion free condition of the metric $de^-+\Omega^-=0$ and the conditions in
\ref{lsgsu4}. It implies that the rotation of $X$ is trivial, i.e.~$e^-\wedge de^-=0$.

The remaining geometric conditions along the transverse directions can be expressed as
\bea
&&2(W_3)_{\bar\a\b\g}[f^2-(g_2+ig_1)^2]+(W_2)_{\bar\a\bar\d_1\bar\d_2} \epsilon^{\bar\d_1\bar\d_2}{}_{\b\g}=0~,
\cr
&&(W_4-2W_5)_{\bar\a}=-\partial_{\bar\a}\log\big[e^{{1\over2}H} {1-2g_2^2-2ig_1g_2\over1-2g_2^2+2ig_1g_2}\big]
\cr
&&(W_1)_{\bar\a_1\bar\a_2\bar\a_3}=i{1-2g_2^2-2i g_1g_2\over 8f^2 g_2^2} \epsilon^\b{}_{\bar\a_1\bar\a_2\bar\a_3}
\partial_\b H
\cr
&&
(W_4)_\a={1\over 8 f^2 g_2^2}\partial_\a H~,
\la{ggsu2}
\eea
where $W_1, W_2, W_3$ and $W_5$ are the Gray-Hervella classes,  see \cite{gray,salamonb}, which can be expressed in terms of  fundamental $SU(4)$ forms $\omega$
and $\chi$ as described in appendix \ref{sun}. Since $W_1$ and $W_2$ do not necessary vanish, one concludes that
the almost complex structure that arises from the metric and $\omega$ in the transverse directions is {\sl not}
integrable.

There is an additional condition that arises from the Killing spinor equations which restricts the functions
$f, g_1, g_2$ that determine the Killing spinors. This is most easily expressed by adapting a coordinate $u$ along $X$, $X={\partial\over\partial u}$,
and introduce coordinates $(u,v, y^I)$ on the spacetime $M$ such that the metric is written as
 \bea
 ds^2=2 e^- e^++\delta_{ij} e^i e^j~,~~~e^-=dv+m_i e^i~,~~~e^+=du+Vdv+n_i e^i~, ~~~e^i=e^i{}_I dy^I~,
 \la{adframe}
 \eea
 where all components depend on $v, y^I$. The Killing spinor equations then imply that
 $f, g_1, g_2$ depend only on the coordinates $y,v$ and the ratio
\bea
\xi(v)={(f+i g_1)^2-g^2_2\over (f-i g_1)^2-g^2_2}~,~~~\xi^*=\xi^{-1}~,
\la{ggsu3}
\eea
depends only on $v$. To summarize, the Killing spinor equations of generic $SU(4)\ltimes\bR^8$-invariant spinors
imply the geometric conditions (\ref{ggsu1}) and (\ref{ggsu2}), and (\ref{ggsu3}).


The conditions that arise from the transverse directions are not particularly illuminating and we have not found a way
to simplify them. It is more straightforward to understand the conditions along the light-cone directions (\ref{ggsu1}).
In particular,  since $X$
has trivial twist there is a coordinate, which we again denote by  $v$,  such that the metric can be written as
\bea
ds^2=2 e^{-{1\over2} H} dv (du+ V dv+ n_i e^i)+ \delta_{ij} e^i e^j~,
\la{simmetr}
\eea
where all components depend on $v,y^I$.
This concludes the investigation of the geometry.

\subsection{Fluxes}

To find the conditions on the fluxes imposed by supersymmetry, we decompose  the forms $\Phi$,  $\Psi$ and ${\cal X}$
in (\ref{mfsu4}) that determine the five-form field strength  into $SU(4)$ representations. Using
the fact that $\Phi$ is anti-self-dual
in the transverse directions,
 one can show
that
\bea
\Phi=\omega\wedge \a+{1\over2} s\bar\wedge {\rm Re}\chi~,
\eea
where $\a$ is a traceless (1,1)-form, $\a_{\b\g}=\a_\b{}^\b=0$, $s$ is a (2,0) and (0,2) symmetric tensor,
$s_{ij}=s_{ji}$, $s_{\a\bar\b}=0$ and $\bar\wedge$ denotes inner derivation\footnote{Let $\pi$ be a k-form, then
$s\bar\wedge\pi={1\over (k-1)!} s^j{}_{i_i} \pi_{j i_2\dots i_k} e^{i_1}\wedge e^{i_2}\wedge\dots \wedge e^{i_k}$.}.
In turn,  one finds that
\bea
\a_{\b\bar\g}={i\over2} \Phi_{\b\bar\g\d}{}^\d~,~~~s_{\bar\a\bar\b}={1\over6}\Phi_{\g_1\g_2\g_3(\bar\a}
\epsilon^{\g_1\g_2\g_3}{}_{\bar\b)}~.
\eea
Similarly using the self-duality of $\Psi$, one can write either
\bea
\Psi={1\over2}  {\rm Re}(p\chi)+{1\over2} w\bar\wedge{\rm Re}\,\chi+ q \omega\wedge\omega+ \hat\Psi^{2,2}~,
\eea
or
\bea
\Psi={1\over2}  {\rm Re}\,(p\, \chi)+\b\wedge \omega+ q \omega\wedge\omega+ \hat\Psi^{2,2}~,
\eea
where
\bea
&&p={1\over 4!}\Psi_{\a_1\dots\a_4} \epsilon^{\a_1\dots \a_4}~,~~~w_{\bar\a\bar\b}=-{1\over6}
\Psi_{\g_1\g_2\g_3[\bar\a} \epsilon^{\g_1\g_2\g_3}{}_{\bar\b]}~,~~~q=-{1\over24} \Psi_\a{}^\a{}_\b{}^\b~,
\cr
&&\b_{\a_1\a_2}={i\over2} \Psi_{\a_1\a_2\b}{}^\b~,~~~i\b_{\a_1\a_2}={1\over2} w_{\bar\g_1\bar\g_2} \epsilon^{\bar\g_1\bar\g_2}{}_{\a_1\a_2}
\eea
and $\hat\Psi^{2,2}$ is a traceless (2,2)-form.

Moreover, the three-form ${\cal X}$ can be written as
\bea
{\cal X}={1\over2} v\bar\wedge {\rm Re}\,\chi+\omega\wedge \g+\hat{\cal X}^{2,1}+\hat{\cal X}^{1,2}~,
\eea
where
\bea
v_{\bar \a}={1\over6}{\cal X}_{\b_1\b_2\b_3}\epsilon_{\bar\a}{}^{\b_1\b_2\b_3}~,~~~ \g_\a={i\over3}{\cal X}_{\a\b}{}^\b~,~~~
\eea
and $\hat{\cal X}^{2,1}$ and $\hat{\cal X}^{1,2}$ are traceless (2,1)- and (1,2)-forms, respectively.

The supersymmetry conditions
imply restrictions on the various irreducible representations of $SU(4)$ that appear in the above decompositions.
In particular, it turns out that by inspecting the conditions in \ref{lsgsu4} one obtains
\bea
\Phi=0~,
\eea
and so
\bea
F=e^-\wedge \Psi+ e^+\wedge e^- \wedge {\cal X}+ {}^*[e^+\wedge e^- \wedge {\cal X}]~.
\eea
The remaining conditions\footnote{We  use  $(\a\cdot \b)_{j_1\dots j_\ell}
={1\over k!} \a_{i_1\dots i_k} \b^{i_1\dots i_k}{}_{j_1\dots j_\ell}$.} give
\bea
&&p={i\over8 f g_2}\big(\partial_-[f^2-(g_2-ig_1)^2]-{i\over8} \nabla_-{\rm Re}\chi\cdot {\rm Im}\chi [f^2-(g_2-ig_1)^2]\big)~,
\cr
&&\b_{\a_1\a_2}=-{i\over 8 f g_2}\big((\nabla_-\omega)_{\a_1\a_2}+{1\over2}[f^2-(g_1-i g_2)^2] (\nabla_-\omega\cdot {\rm Re}\,\chi)_{\a_1\a_2}\big)~,
\cr
&&q={1\over 24 f g_2}\big(2 g_2\partial_-g_1-2 g_1\partial_-g_2+{1\over8} \nabla_-{\rm Re}\chi\cdot {\rm Im} \chi\big)~,
\cr
&&v_\a=-{i\over8} [f^2-(g_2+i g_1)^2] \partial_\a \log{1+2 f g_2\over 1-2f g_2}~,
\cr
&&{\cal X}^{2,1}+{\cal X}^{1,2}={1\over 8 f g_2}\omega\wedge (de^-)_{-i} e^i-{f g_2\over 2} (d\omega^{2,1}+ d\omega^{1,2})~.
\eea
Observe that $\Psi^{2,2}$ is not restricted by the Killing spinor equations.
One can substitute the above expressions into the formula for $F$. We shall not do this here because it does not lead to
a simplification for  the expression of $F$. However,  as we shall show, $F$ is simplified in some special cases.

\subsection{Special cases}

A large class of backgrounds consists of those for which the transverse  metric is independent of $v$. Using the
torsion free condition for the frame $(e^-, e^+, e^i)$, one finds that
\bea
\nabla_-\omega^{2,0}=\nabla_-\chi^{4,0}=0~,
\la{vanvan}
\eea
provided that\footnote{With this notation we mean that there are forms $\l=\l_i e^i$ and $\mu={1\over2}\mu_{ij} e^i\wedge e^j$
such that $de^+=e^-\wedge \l+\mu$, where $\mu$ is  (1,1) and  traceless.}  $de^+\in \mathfrak{su}(4)\oplus\bR^8$. Assuming also that
 Killing spinors
are taken to be independent of $v$ as well, we have that
\bea
p=q=\b=0~.
\eea
In such a case, the  flux can be written as
\bea
F&=&  {1\over2} e^+\wedge e^-\wedge v\bar\wedge {\rm Re}\,\chi+{1\over 8 f g_2}e^+\wedge \omega\wedge de^-
\cr&&- {f g_2\over 2} e^+\wedge e^-\wedge  (d\omega^{2,1}+ d\omega^{1,2})
+ {}^*[e^+\wedge e^-\wedge {\cal X}]+e^-\wedge \hat\Psi^{2,2}~.
\eea
To construct solutions in this case, one must find almost Hermitian manifolds with an $SU(4)$
structure which satisfy the conditions  (\ref{ggsu2}), and then write the metric as (\ref{simmetr}).
One must also impose the closure of $F$, $dF=0$ and the $E_{--}=0$ component of the Einstein equations.

\newsection{Backgrounds with pure $SU(4)\ltimes\bR^8$ invariant Killing spinors }
\la{spsu4}
\subsection{Geometry and fluxes}

In the solution of the Killing spinor equations with $SU(4)\ltimes\bR^8$ invariant Killing spinors, a special case arises
whenever the Killing spinor is pure  \cite{ugjggpa}. In particular, the Killing spinors (\ref{kssu4}) are pure if one sets
\bea
g_1=0~,~~~f=\pm g_2={h\over2}~.
\eea
 We  consider the case $f+g_2=0$, the investigation
of the other case is similar. The solution of the linear system can be found in appendix \ref{lspsu4}.
Here we shall investigate  the conditions on the geometry and write the fluxes in a closed form.

It is convenient to investigate the geometry in the gauge $h=1$. First write the metric and fluxes
as in (\ref{mfsu4}) using the pseudo-hermitian frame $(e^-, e^+, e^\a, e^{\bar\a})$. Next consider the form bi-linears (\ref{bisu4})
and observe that the conditions on the geometry that involve light-cone directions can be written as
\bea
{\cal L}_X g={\cal L}_X( e^-\wedge \omega)={\cal L}_X(e^-\wedge \chi)=0~,~~~de^-\in \mathfrak{su}(4)\oplus \bR^8~,
\la{gpsu1}
\eea
i.e.~$X=e_+$ is Killing vector field and preserves the $SU(4)\ltimes\bR^8$ structure. Unlike the
generic $SU(4)\ltimes\bR^8$ case, $e^-\wedge de^-\not=0$.
The remaining geometric conditions along the transverse directions are
\bea
\Omega_{\a,\b\g}=0~,~~~\Omega_{\bar\a,\b}{}^\b+\Omega_{\b,\bar\a}{}^\b=0~,~~~\Omega_{\bar\a,\b}{}^\b=-\Omega_{-,+\bar\a}~,
\eea
which can be recast in terms of the $SU(4)$ structures, see  appendix  \ref{sun}, as
\bea
W_1=W_2=0~,~~~W_4=W_5~,~~~(W_4)_i= (de^-)_{-i}~.
\la{gpsu2}
\eea
The vanishing of $W_1, W_2$ can be interpreted as integrability of the almost complex structure
along the transverse directions. In particular, one can show using the torsion free conditions of the frame, that
$(e^-, e^{\a})$ span an integrable distribution of co-dimension 5. As we have seen, this is unlike what happens in
the generic $SU(4)\ltimes\bR^8$ backgrounds. The conditions
(\ref{gpsu1}) and (\ref{gpsu2}) constitute the full set of restrictions that supersymmetry imposes on the geometry of  spacetime.

Next let us turn to the flux $F$. Using (\ref{mfsu4}) and the results of appendix \ref{lspsu4}
  $F$  can be written, after some work, as
\bea
F&=&-{1\over4} e^+\wedge d (e^-\wedge \omega)+{}^*[e^+\wedge e^-\wedge {\cal X}]
+{i\over4} e^-\wedge \nabla_-\omega^{2,0}\wedge \omega-{i\over4} e^-\wedge \nabla_-\omega^{0,2}\wedge \omega
\cr
&&
-{1\over 2^2\cdot 4!} e^-\wedge \omega\wedge \omega [\nabla_-{\rm Re}\chi\cdot {\rm Im} \chi]
+e^-\wedge \hat\Psi^{2,2}~,
\label{purespin5form}
\eea
where $\a\cdot \b={1\over k!} \a_{i_1\dots i_k} \b^{i_1\dots i_k}$. It is clear that all the components of $F$
are determined in terms of the geometry apart from those of $\hat\Psi^{2,2}$ which is not restricted by
the Killing spinor equations. In this case, $\Phi$ may not vanish.

One can adapt coordinates along the Killing vector field $X$, $X=\partial/\partial u$, and write the
metric as (\ref{adframe}). However unlike the generic case no further simplification is possible because
the twist of $X$ may not be trivial, i.e.~$e^-\wedge de^-\not=0$. If one imposes $e^-\wedge de^-=0$, one finds
additional restrictions on the geometry that are {\it not} implied by the Killing spinor equations.

\subsection{Special cases and examples}

\subsubsection{Special cases}
As in the class of generic backgrounds,  one can take the transverse metric to be independent  of $v$ and impose
$de^+\in \mathfrak{su}(4)\oplus \bR^8$ to find (\ref{vanvan}). In such cases, $F$ can be written as
\bea
F=-{1\over4} e^+\wedge d (e^-\wedge \omega)+{}^*[e^+\wedge e^-\wedge {\cal X}]
+e^-\wedge \hat\Psi^{2,2}~.
\la{sflux}
\eea
Further simplification of the metric and fluxes occurs when the rotation of $X$ is trivial, i.e.~$e^-\wedge de^-=0$. This is equivalent to requiring that there is an one-form $\l$ such that
$de^-=\l\wedge e^-$ or equivalently $de^-\in \bR^8$. Then the Frobenius theorem implies that  there is a
function $H=H(y,v)$  such that
$e^-= e^{H(y,v)} dv$ for some coordinate $v$ which is related to that denoted with the same symbol in (\ref{adframe})
by a coordinate transformation.
The metric takes the form
\bea
ds^2= 2e^H dv (du+ Vdv+ n_I dy^I)+g_{IJ}(y) dy^I dy^J~.
\la{smetr}
\eea
A large class of known backgrounds have metric and fluxes given by (\ref{smetr}) and  (\ref{sflux}), respectively.
These include the D3-brane and intersecting D3-brane configurations as we shall see below.

Moreover, the geometric conditions are also simplified. In particular one finds that
the geometric conditions in (\ref{gpsu2}) become
\bea
W_1=W_2=0~,~~~(W_4)_I=(W_5)_I=-\partial_I H~.
\eea
The spacetime in this case can be reconstructed from an eight-dimensional  Hermitian manifold with a $SU(4)$ structure
for which the $W_4$ and $W_5$ classes satisfy the conditions above. Assuming that $H$ depends only on $y$, $H=H(y)$, and setting
$W_1=W_2=0$ in  (\ref{exexsu4}) which expresses  the exterior derivatives of $\omega$ and $\chi$ in terms
of the $W$ classes,  we find that the remaining geometric conditions can be rewritten  as
\bea
d( e^H \omega^3)=d (e^H \chi)=0~.
\la{clocon}
\eea
Observe that the rescaled forms $\omega_0=e^{{1\over3} H} \omega$ and $\chi_0=e^H \chi$ for $dH\not=0$ are not canonically normalized,
so such Hermitian manifolds do not contain a Calabi-Yau in their conformal class.
The expressions for the metric (\ref{smetr}) and the flux (\ref{sflux}),  and the conditions (\ref{clocon})
are the full content of the Killing spinor equations for the case that the transverse metric and $H$ are independent of $v$.

Of course additional conditions are imposed on the backgrounds from the Bianchi identity of $F$, $dF=0$ and the vanishing
of the $E_{--}$ component of the Einstein equations, $E_{--}=0$.
In what follows, we consider  examples  which admit a null Killing vector with a trivial and a non-trivial twist.

\subsubsection{D3-branes and intersecting branes}

The D3-brane \cite{d3brane} and its  intersections \cite{paptown} are
examples of backgrounds with pure $SU(4)\ltimes\bR^8$-invariant Killing spinors.
The metric of the former is
\bea
ds^2= h^{-{1\over2}} ds^2(\bR^{3,1})+ h^{{1\over2}} ds^2(\bR^6)~,
\eea
where $h$ is a (multi-centred)  harmonic function of $\bR^6$. ($h$ should be distinguished
from the function $h$ that multiplies the Killing spinor which we have set equal to 1.)
The background preserves 16 supersymmetries.
Clearly, this metric is of the form (\ref{smetr}). To rewrite the metric as (\ref{adframe}),
set
\bea
ds^2=2 h^{-{1\over2}} dx dz+  h^{-{1\over2}} ds^2(\bR^2)+ h^{{1\over2}} ds^2(\bR^6)
\eea
and change coordinates as
\bea
u=x~,~~~v=h^{-{1\over2}} z
\eea
to find
\bea
ds^2=2 e^- e^++  h^{-{1\over2}} ds^2(\bR^2)+ h^{{1\over2}} ds^2(\bR^6)~,~~~e^-=dv+{1\over2} v\, d\log h~,~~~e^+=du~.
\eea
Observe that the pure spinor $1$ satisfies the projection condition that arises in the  D3-brane Killing spinor equations.
It is then easy to find that
\bea
\omega=h^{-{1\over2}} \omega(\bR^2)+ h^{1\over2} \omega(\bR^6)
\eea
where $\omega(\bR^2)$ and $\omega(\bR^6)$ are the constant K\"ahler forms on $\bR^2$ and $\bR^6$, respectively.
Substituting all these into the flux and taking $\hat\Psi^{2,2}=0$, one can easily show that
\bea
F&=&-{1\over4} h^{-{3\over2}} du\wedge dv\wedge \omega(\bR^2)\wedge dh+ {}^*[e^+\wedge e^-\wedge {\cal X}]
\cr
&=&
{1\over4}  dx\wedge dz\wedge \omega(\bR^2)\wedge dh^{-1}+ {}^*[e^+\wedge e^-\wedge {\cal X}]~.
\eea
This is precisely the five-form flux of the D3-brane. Moreover observe that the space transverse
to the lightcone directions is Hermitian,
the closure conditions (\ref{clocon}) are satisfied and the rotation of $X$ is trivial $e^-\wedge de^-=0$.
The latter also follows from a direct inspection of the D3-brane metric.  One consequence of
these conditions is that $H_3\times S^5$, which
is the near-horizon geometry of the D3-brane restricted on the transverse directions to the lightcone,
admits a relatively balanced $SU(4)$  Hermitian structure.

One can also show that this class includes the delocalized D3-brane intersections. We shall demonstrate this
for the configuration of two intersecting D3-branes at a string \cite{paptown}. The computation is straightforward for the
remaining cases. The metric  is \cite{tseytlinh}
\bea
ds^2=2 (h_1 h_2)^{-{1\over2}} dx dz+ h_1^{-{1\over2}}h_2^{{1\over2}} ds_1^2(\bR^2)+
h_1^{{1\over2}}h_2^{-{1\over2}} ds_2^2(\bR^2)+h_1^{{1\over2}}h_2^{{1\over2}} ds^2(\bR^4)
\eea
where $h_1, h_2$ are (multi-centred) harmonic functions of the transverse space $\bR^4$. Changing coordinates
as
\bea
u=x~,~~~v=(h_1 h_2)^{-{1\over2}} z
\eea
the metric can be written in the standard light-cone form with
\bea
e^+=du~,~~~e^-=dv+{1\over2} v\, d\log (h_1 h_2)=(h_1 h_2)^{-{1\over2}} dz
\eea
The hermitian form
can be chosen as
\bea
\omega=h_1^{-{1\over2}}h_2^{{1\over2}} \omega_1(\bR^2)+ h_1^{{1\over2}}h_2^{-{1\over2}} \omega_2(\bR^2)
+h_1^{{1\over2}}h_2^{{1\over2}} \omega(\bR^4)~.
\eea
Substituting this into the the expression for the flux, one finds that
\bea
F={1\over4}  dx\wedge dz\wedge \omega_1(\bR^2)\wedge dh_1^{-1}
+{1\over4}  dx\wedge dz\wedge \omega_2(\bR^2)\wedge dh_2^{-1}+{}^*[e^+\wedge e^-\wedge {\cal X}] ~ .
\eea
Similarly observe that the transverse
space is Hermitian,
the closure conditions (\ref{clocon}) are satisfied and the rotation of $X$ is trivial $e^-\wedge de^-=0$.

One can easily extend the above results to include D3-brane configurations with a null rotation by taking $n\not=0$
and superposed with a pp-wave $V\not=0$. One can also allow $\hat\Psi^{2,2}\not=0$ which may lead to resolved
D3-brane configurations.

\subsubsection{Uplifted Five-Dimensional Solutions}

Solutions of minimal gauged five-dimensional supergravity which admit a timelike Killing vector, ${\partial \over \partial t}$,
associated with  a Killing spinor
have spacetime geometry \cite{gutgaunfive}
\be
ds_5^2 = -{\cal{F}}^2(dt + \Psi)^2+{\cal{F}}^{-1}ds^2_N ~,
\ee
where   ${\cal{F}}$ is a function and
$\Psi=\Psi_m dx^m$ is a 1-form, $ds^2_N=h_{mn} dx^m dx^n$ is a metric on a K\"ahler 4-manifold $N$, and  $(t, x^m)$, $m=1,2,3,4$, are spacetime coordinates.
The components of the metric depend only on $x^n$. In addition, the one-form gauge potential is
\be
A = {\sqrt{3} \over 2} {\cal{F}} (dt + \Psi) + {\ell \over 2 \sqrt{3}} {\cal{P}}~,
\ee
where $\ell$ is constant and ${\cal{R}}_N = d {\cal{P}}$ is the Ricci form of the K\"ahler manifold $N$.
The function ${\cal{F}}$ is determined in terms of the Ricci scalar $R_N$ of $N$ via
\be
{\cal{F}} = -{24 \over \ell^2 R_N}~.
\ee
Setting
\be
\cF d \Psi = G^+ + G^-~,
\ee
where $G^+$, $G^-$ are self and anti-self-dual 2-forms on $N$,
the Ricci form ${\cal{R}}_N$ is constrained by
\be
{\cal{R}}_N = -{2 \over \ell} G^+ -{6 \over \cF \ell^2} J_N~,
\ee
where $J_N$ is the K\"ahler form of $N$.

The uplifted metric is given by \cite{chamemp}

\bea
ds_{10}^2 &=& ds_5^2 + \ell^2 \big( d \alpha^2+ \cos^2 \alpha d \beta^2
+ \sin^2 \alpha \cos^2 \alpha (d \xi_1-\sin^2 \beta d \xi_2 - \cos^2 \beta d \xi_3)^2
\cr
&+& \cos^2 \alpha \sin^2 \beta \cos^2 \beta (d \xi_2-d \xi_3)^2\big)
\cr
&+&(-{2 \over \sqrt{3}}A - \ell \sin^2 \alpha d \xi_1
-\ell \cos^2 \alpha \big(\sin^2 \beta d \xi_2 + \cos^2 \beta d \xi_3)\big)^2~.
\eea
It is convenient to define $\chi_1 =\xi_1 -{1 \over 2}(\xi_2+\xi_3)$,
$\chi_2 = \xi_1 +{1 \over 2}(\xi_2+\xi_3)$, $\phi = {1 \over 2}(\xi_2-\xi_3)$ and rewrite the metric as
\bea
ds^2 &=& {2\ell {\cal{F}} \over 3} \big(dt+\Psi + {\ell \over 4 {\cal{F}}}  d \chi_2+
{\ell \over 6 {\cal{F}}}({\cal{P}}+{\cal{Q}})\big)
\big ({3 \over 2} d \chi_2 +{\cal{P}}+{\cal{Q}}\big )
\cr
& &+ {\cal{F}}^{-1}h_{mn} dx^m dx^n + \ell^2 ds_{CP^2}^2~,
\eea
where
\bea
\label{cp2met}
 ds_{CP^2}^2 &=& d \alpha^2+ \cos^2 \alpha d \beta^2
 + \sin^2 \alpha \cos^2 \alpha (d \chi_1+ (\cos^2 \beta-\sin^2 \beta)d \phi)^2
 \cr
 &+&4 \cos^2 \alpha \sin^2 \beta \cos^2 \beta d \phi^2~,
 \eea
 is the K\"ahler-Einstein metric on $CP^2$ which has constant holomorphic sectional curvature.
 ${\cal{Q}}$ is the potential for the Ricci form of the metric ({\ref{cp2met}}), where
 \be
 {\cal{Q}} = 3 \cos^2 \alpha (\sin^2 \beta-\cos^2 \beta)d \phi +{3 \over 2}(\sin^2 \alpha-
 \cos^2 \alpha)d \chi_1~,
 \ee
 and the K\"ahler form of $CP^2$ is  $J_{CP^2} = {1 \over 6} d {\cal{Q}}$.
The Ricci scalar of ({\ref{cp2met}})  is $R_{CP^2} = 24$.
Note that in the uplifted solution, the Killing vector ${\partial \over \partial t}$ is null.

In order to write the uplifted solution as a pure spinor $SU(4)\ltimes \bR^8$ background, we set
$ds^2=2 e^- e^++ds^2_8$, where
\bea
e^+ &=&(dt+\Psi + {\ell \over 4 {\cal{F}}}  d \chi_2+
{\ell \over 6 {\cal{F}}}({\cal{P}}+{\cal{Q}}))~,
\cr
e^- &=& {\ell {\cal{F}} \over 3}
({3 \over 2} d \chi_2 +{\cal{P}}+{\cal{Q}})~,
\cr
ds^2_8 &=&  {\cal{F}}^{-1}ds^2_N + \ell^2 ds_{CP^2}^2~.
\eea

The calculation we present below
can also be carried out if $CP^2$ is replaced with any other four-dimensional K\"ahler-Einstein manifold $E$
such that $R_E=24$. However, we shall continue the analysis using $CP^2$.

The complex structure along the transverse directions to the lightcone is identified as the direct sum
of the complex structure on $CP^2$ together with the complex structure of
the K\"ahler base manifold $N$ of the 5-dimensional solution. It is clear that the null Killing vector field
$\partial/\partial t$ of the ten-dimensional solution has non-trivial twist.

It will be convenient to split the $SU(4)$ indices $\alpha$
into $\alpha=(a,\mu)$ for $a,b =1,2$ and $\mu,\nu=3,4$
and choose a Hermitian frame along the transverse directions as
\bea
e^a &=&  e^{ik_1 \chi_2}  \cF^{-{1 \over 2}} {\hat{e}}^a
\cr
e^\mu &=& e^{ik_2 \chi_2} \ell {\hat{e}}^\mu
\eea
for real constants $k_1, k_2$ to be fixed, where
${\hat{e}}^a$ is a Hermitian frame of the K\"ahler base manifold $N$,
$
ds^2_N = 2 \delta_{{a} {\bar{{b}}}}
{\hat{e}}^a {\hat{e}}^{\bar{b}}
$, $J_N = -i \delta_{{a} {\bar{{b}}}}
{\hat{e}}^a \wedge {\hat{e}}^{\bar{b}}$~,
such that
\be
\cL_{\partial \over \partial t} {\hat{e}}^a = \cL_{\partial \over \partial \chi_1} {\hat{e}}^a = \cL_{\partial \over \partial \chi_2} {\hat{e}}^a
= \cL_{\partial \over \partial \alpha} {\hat{e}}^a
= \cL_{\partial \over \partial \beta} {\hat{e}}^a
= \cL_{\partial \over \partial \phi} {\hat{e}}^a =0~.
\ee
We also set
\bea
{\hat{e}}^3 &=& {1 \over \sqrt{2}}
\big(d \alpha -i \sin \alpha \cos \alpha (d \chi_1+(\cos^2 \beta - \sin^2 \beta)d \phi) \big)~,
\cr
{\hat{e}}^4 &=& {1 \over \sqrt{2}}
\big( \cos \alpha d \beta -2i \cos \alpha \sin \beta \cos \beta d \phi \big)~.
\eea
In addition, we have
$
ds^2_{CP^2} = 2 \delta_{{\mu} {\bar{{\nu}}}}\,
{\hat{e}}^\mu {\hat{e}}^{\bar{\nu}}
$
and $
 J_{CP^2} = -i \delta_{{\mu} {\bar{{\nu}}}}\,
{\hat{e}}^\mu \wedge {\hat{e}}^{\bar{\nu}}$.
 It is also convenient to write
$
\omega= \omega_N + \omega_{CP^2}
$,
where
\be
\omega_N = \cF^{-1} J_N , \qquad \omega_{CP^2} = \ell^2 {\tilde{J}}_{CP^2}~.
\ee
Using the expression for the spin connection which we give in
\ref{spincon},  it is straightforward to show that
 the constraints on the geometry imposed by supersymmetry are satisfied provided that
 \be
 k_1 + k_2 = -{3 \over 4}~.
 \ee

 Next, note that the formula for the uplifted five-form is
 \cite{chamemp}
 \be
 F = {1 \over 4}(1+*) \big(-{4 \over \ell}
 dvol_5  +{\ell^2 \over \sqrt{3}}
 \sum_{i=1}^3 d(\mu_i^2) \wedge d \xi_i \wedge *_{(5)} dA\big)
 \ee
where $\mu_1=\sin \alpha$, $\mu_2= \cos \alpha \sin \beta$, $\mu_3= \cos \alpha \cos \beta$ and
 \be
 d{\rm vol}_5 = -{1 \over 2} \cF^{-2} e^0 \wedge J_N \wedge J_N~,
 \ee
 with $e^0 = \cF(dt+\Psi)$. In addition,
 $*_{(5)}$ denotes the 5-dimensional Hodge dual taken with
 respect to $d{\rm vol}_5 $ and $*$ denotes the standard
 Hodge star operation of $M$ whose volume form can be rewritten as
 $d{\rm vol}_M={1 \over 4} e^+ \wedge e^- \wedge \omega_N \wedge \omega_N
 \wedge \omega_{CP^2} \wedge \omega_{CP^2}$.
 Note the change in normalization for the 5-form
 when comparing with the Killing spinor equation in
 \cite{jgjg}.
In order to simplify this expression, note that
 \be
  \sum_{i=1}^3 d(\mu_i^2) \wedge d \xi_i  = 2 J_{CP^2}~.
 \ee

 We then find the uplifted five-form is given by
\bea
 F &=& {1 \over 4} \ell^2 \cF^{-1} e^+ \wedge e^- \wedge J_{CP^2}  \wedge d \cF +*({1 \over 4} \ell^2 \cF^{-1} e^+ \wedge e^- \wedge
 J_{CP^2}  \wedge d \cF)
  \cr
 &+& e^+ \wedge \big({1 \over 2} \cF^{-1}\ell^{-1} J_N \wedge J_N
 -{1 \over 2}\ell^3 \cF  J_{CP^2} \wedge   J_{CP^2}
 +{1 \over 6} \cF \ell^2   J_{CP^2} \wedge G^+ \big)
 \cr
 &+& e^- \wedge (-{1 \over 4}\ell^{-1} \cF^{-1} \omega \wedge \omega
 +{1 \over 4} \ell^2 \cF^{-1}  J_{CP^2} \wedge G^-)
  \eea

  In comparing this expression with the 5-form obtained from the
  pure spinor classification, it is straightforward to see that
\bea
 -{1 \over 4} e^+ \wedge d(e^- \wedge \omega) &=&
   e^+ \wedge \big({1 \over 2} \cF^{-1}\ell^{-1} J_N \wedge J_N
 -{1 \over 2}\ell^3 \cF  J_{CP^2} \wedge   J_{CP^2}
 +{1 \over 6} \cF \ell^2   J_{CP^2} \wedge G^+ \big)
 \cr
 &+& {1 \over 4} \ell^2 \cF^{-1} e^+ \wedge e^- \wedge J_{CP^2}  \wedge d \cF~.
\eea
Also note that
\bea
{i \over 4} \nabla_- \omega^{2,0} \wedge \omega -{i \over 4} \nabla_-\omega^{0,2} \wedge \omega -{1 \over 4\cdot 4!}
(\nabla_- {\rm Re} \chi)\cdot {\rm Im} \chi\, \omega \wedge \omega =
\cr
-{1 \over 4}\ell^{-1} \cF^{-1} \omega \wedge \omega
+{1 \over 4} \ell^2 \cF^{-1}  J_{CP^2} \wedge G^-
\cr
 +{i \over 24} \cF^{-1} (G^-)_a{}^a (\omega_N
 \wedge \omega_N - \omega_N \wedge \omega_{CP^2}
+\omega_{CP^2} \wedge \omega_{CP^2})~.
 \eea

However, $\omega_N
 \wedge \omega_N - \omega_N \wedge \omega_{CP^2}
+\omega_{CP^2} \wedge \omega_{CP^2}$
is a traceless (2,2) form. We therefore set
\be
{\hat{\Psi}}^{2,2} = -{i \over 24} \cF^{-1} (G^-)_a{}^a (\omega_N
 \wedge \omega_N - \omega_N \wedge \omega_{CP^2}
+\omega_{CP^2} \wedge \omega_{CP^2})
\ee
Hence, the formula for the uplifted five-form matches
the expression for the five-form given in ({\ref{purespin5form}}). It is remarkable
that in establishing the equality of the uplifted five-form with that in ({\ref{purespin5form}}), we have
seen that all components
of the latter give a non-trivial contribution.

It remains to consider the Bianchi identity; note that one can
rewrite  $F$ as

\bea
F &=& {1\over 4} d(e^+ \wedge e^- \wedge \omega) -{\ell^2 \over 4} J_{CP^2} \wedge \star_N d \cF^{-1}
\cr
&+&\cF^{-1} e^- \wedge \big( {\ell^3 \over 12} J_{CP^2} \wedge {\cal{R}}_N -{1 \over 4} J_N \wedge d \Psi
-{\ell^3 \over 2} J_{CP^2} \wedge J_{CP^2} \big)
\eea
where $\star_N$ denotes the Hodge dual taken on the K\"ahler base
with metric $h$ and volume form $-{1 \over 2}J_N \wedge J_N$.
Hence
\be
dF = J_{CP^2} \wedge \big(-{\ell^2 \over 4} d\star_N d \cF^{-1} +{\ell^4 \over 36} {\cal{R}}_N \wedge {\cal{R}}_N
-{\ell \over 2} J_N \wedge d \Psi \big) \ .
\ee
So the Bianchi identity implies that
\be
-d \star_N d \cF^{-1} + {4 \over 9} G^+ \wedge G^+ +4 \ell^{-2} \cF^{-2} J_N \wedge J_N -2 \ell^{-1} \cF^{-1} J_N \wedge G^-=0 \ .
\ee
As expected, this constraint is equivalent to that implied by the five-dimensional
gauge field equations obtained in  \cite{gutgaunfive}. Observe that as $\cF$ is inversely proportional to
the Ricci scalar $R_N$, this constraint is a highly non-linear constraint on the geometry of the
K\"ahler base space $N$.

The $AdS_5$ black holes found in {\cite{hjg}} belong to the class of five-dimensional backgrounds which we have
uplifted to ten dimensions. In fact, it was shown in \cite{jgjg}
that the uplifted black hole solutions preserve exactly $1 /16$
of the supersymmetry in IIB supergravity. Hence it follows that
the Killing spinors of the uplifted black hole solution are $(1, i1)$ of
the pure spinor $SU(4)\ltimes\bR^8$ backgrounds. This can also be seen by an inspection of the conditions
given in \cite{jgjg}.

\newsection{$Spin(7)\ltimes\bR^8$ backgrounds}
\la{sspin7}
As in previous cases, it is convenient to carry out the analysis without loss of generality in the gauge
$f^2+g^2=1$. A direct inspection of the geometric conditions in appendix \ref{lsspin7} reveals that the holonomy
of the Levi-Civita connection, $\nabla$,  of the spacetime is contained in $Spin(7)\ltimes\bR^8$,
\bea
{\rm hol}(\nabla)\subseteq Spin(7)\ltimes\bR^8~.
\eea
This is equivalent to requiring that the forms
\bea
e^-,~~~ e^-\wedge\phi~,
\eea
 are
$\nabla$-parallel,  where $\phi$ is the fundamental
$Spin(7)$ self-dual four-form given by $\phi={\rm Re}\, \chi-{1\over2}\omega\wedge \omega$.
In particular, the null vector field $X$ is Killing and twist free.
Adapting
coordinates along the parallel vector field, $X=e_+=\partial/\partial u$, one finds that the spacetime metric can be written as
\bea
ds^2=2 dv (du + V dv+ n_I dy^I)+ \gamma_{IJ} dy^I dy^J~,
\eea
where the components depend on the coordinates $v, y^I$.
It remains to specify the fluxes. One can show  using the results of appendix \ref{lsspin7} that
\bea
F={i\over 14} \partial_-\log({f+ig\over f-ig})\,  e^-\wedge\phi+e^-\wedge \Psi_{{\bf 27}}~.
\eea
To derive this, we first remark that $F=e^-\wedge \Psi$, where $\Psi$ is a self-dual four-form
in the eight-transverse directions to the light-cone. Then we use the decomposition of the four-forms
in $\bR^8$ under $Spin(7)$ representations as
$\Lambda^4(\bR^8)=\Lambda^{4+}(\bR^8)\oplus \Lambda^{4-}(\bR^8)$, $\Lambda^{4+}(\bR^8)=\Lambda_{{\bf 1}}\oplus
\Lambda_{{\bf 7}}\oplus \Lambda_{{\bf 27}}$ and $\Lambda^{4-}(\bR^8)=\Lambda_{{\bf 35}}$. It turns out that the
Killing spinor equations imply that the
 $\Psi_{{\bf 7}}$  component of $\Psi$ vanishes,  $\Psi_{{\bf 7}}=0$. The component $\Psi_{{\bf 27}}$ is not restricted by the Killing
 spinor equations.
The Killing spinors are
\bea
\epsilon=(f+ig)(v) (1+e_{1234})~,
\eea
i.e.~the functions $f,g$ depend only on $v$. It is clear from the above that the spacetime is a pp-wave with rotation
propagating on an eight-dimensional manifold which has holonomy $Spin(7)$. The metric of the transverse space depends
on the wave-profile coordinate $v$.

Imposing the Bianchi identity for $F$ one finds that $F$ is closed iff $d\Psi_{{\bf 27}}=0$ up to forms of the type $e^-\wedge \mu$.
The only equation that remains to be imposed to find solutions is the vanishing of the $E_{--}$ component
of the Einstein equations, $E_{--}=0$. This equation can be easily recovered from that  in  \cite{iibsyst}
by setting the one-form and three-form field strengths to zero.

\newsection{$G_2$ backgrounds}
\la{sg2}

\subsection{Geometry}

We have presented the linear system for $G_2$ backgrounds and its solution in appendix \ref{lsg2}.
In particular, we have given the solution  in $SU(3)\subset G_2$
representations.   Here, we shall investigate the consequences
that the supersymmetry conditions have on the geometry and fluxes of the theory.
For this, we introduce a  frame\footnote{This is a real frame
and different from the pseudo-Hermitian frame of appendix \ref{lsg2} which we have used to solve the linear system.}
 $(e^+, e^-, e^1, e^i)$, $i=2,3,4,6,7,8,9$, where $(e^+, e^-, e^1)$ span the
trivial subbundle of $TM$ and $(e^i)$ the transverse directions of the spacetime.   The metric and fluxes written in this frame  are
\bea
&&ds^2=2 e^- e^++ (e^1)^2+ \delta_{ij} e^i e^j~,~~~
\cr
&&F= e^+\wedge \Phi+ e^-\wedge \Psi+ e^+\wedge e^-\wedge {\cal X}+{}^*[e^+\wedge e^-\wedge {\cal X}]~,
\la{mfg2}
\eea
where $\Phi$ is a anti-self dual, and $\Psi$ is a self-dual, four-form in the eight directions transverse to
the light-cone directions, and ${\cal X}$ is a three-form. The expression for the flux is similar to that
which we have given for $F$  in (\ref{mfsu4}) for the $SU(4)\ltimes\bR^8$ backgrounds.

Choosing a basis $( 1+e_{1234}, e_{15}+e_{2345})$ in the space of $G_2$-invariant spinors, one can show that
a basis in the space of spinor bilinears is
\bea
e^0~,~~~e^1~,~~~e^5~,~~~\varphi~,
\eea
where
\bea
\label{g2canonform}
\varphi={\rm Re}[(e^2+ie^7)\wedge (e^3+ie^8)\wedge (e^4+ie^9)]- e^6\wedge (e^2\wedge e^7+e^3\wedge e^8+e^4\wedge e^9)
\eea
 is the fundamental $G_2$ three-form, and $e^\pm=(1/\sqrt 2) (\pm e^0+ e^5)$.
It is clear that $T^*M= I^3\oplus {\cal T}$ where $I^3$ is a trivial bundle of rank 3
 and ${\cal T}$ are the remaining ``transverse'' directions of the co-tangent bundle of the spacetime.

The geometric conditions of appendix \ref{lsg2} can be expressed in
 $G_2$ representations as
 \bea
 &&{\cal L}_V g=0~,~~~d(f^2 e^1)= d(f^2 e^5)=0~,~~~(Z_+)_i=-{1\over12} [(Y_+)_i-(Y_-)_i]~,
 \cr
&&(Y_1)_i=-4 (\nabla_i e_-)_+~,~~~
(Y_+)_i=-4 (\nabla_i e_+)_1~,~~~(Y_-)_i=4 (\nabla_i e_-)_1~, ~~~
\cr
&&
\delta^{ij} (\nabla_i e_+)_{j}=(\nabla_1 e_+)_1~,~~~\delta^{ij} (\nabla_i e_1)_{j}=-{1\over2}[(\nabla_- e_-)_1+
(\nabla_+ e_+)_1]~,~~~
\cr
&&X_1=2[(\nabla_+ e_1)_+-(\nabla_- e_1)_-]~,~~~X_2=0~,~~~  X_4=\theta_\varphi=0~,
\la{g2gcona1}
\eea
where the one-form associated with $V$ is $\kappa=f^2 e^0$,
\bea
(Y_r)_i={1\over6}\nabla_r\varphi_{jkl}\, \star\varphi^{jkl}{}_i~,
~~~(Z_r)_i={1\over6} (d e_r)_{kl} \varphi^{kl}{}_i~,~~~r=-,+,1~,~~~
 \eea
and $X_1, X_2, X_3$ and $X_4$ have been defined in appendix \ref{g2}. The $\star$ Hodge duality operation
is with respect to the ``transverse'' volume form $d{\rm vol}=e^2\wedge e^3\wedge e^4\wedge e^6\wedge \dots\wedge e^9$.
Clearly $V$ is a time-like Killing vector field as
may have been expected from the results of \cite{ugjggpa, ugjggpb}. The class $X_3$, which is associated with the traceless symmetric representation
of $G_2$,
is not restricted by the supersymmetry conditions.

To show that $X_2=0$, one has to compute
 \bea
 \Pi_{{\bf 14}}(\star \tilde d \star \varphi) \ ,
 \eea
 and demonstrate that it vanishes using the conditions, where ${\tilde{d}}$ denotes the
 exterior derivative restricted to the transverse directions, and $\Pi_{{\bf 14}}$ is the projection
 to the 14-dimensional representation in the decomposition of $\Lambda^2(\bR^8)=\Lambda_{{\bf 7}}\oplus \Lambda_{{\bf 14}}$
 in $G_2$ irreducible representations, $\Lambda_{{\bf 14}}=\mathfrak{g}_2$. In particular,
\bea
(\Pi_{{\bf 14}} \gamma)_{ij}= {4\over6} ({1\over4} \star\varphi^{kl}{}_{ij} \gamma_{kl}
\,
+\gamma_{ij})~,~~~\g\in\Lambda^2(\bR^8)~.
 \eea
 Moreover, one can immediately see using
 (\ref{cocal}) that
 \be
  \tilde d \star \varphi =0 \ .
  \ee
  Thus the transverse directions are {\it co-symplectic} or {\it co-calibrated}.
  Moreover, it turns out that using the conditions (\ref{g2gcona1}), one can
 show that ${\cal L}_V\varphi=0$.

 It is sometimes more convenient to re-express the conditions on the geometry as
 \bea
&&{\cal L}_V g=0~,~~~d(f^2 e^1)= d(f^2 e^5)=0~,~~~{\cal L}_V\varphi=0~,
 \cr
&&(Y_1)_i=-4 (\nabla_i e_-)_+~,~~~
(Y_+)_i=-4 (\nabla_i e_+)_1~,~~~(Y_-)_i=4 (\nabla_i e_-)_1~, ~~~
\cr
&&
\delta^{ij} (\nabla_i e_+)_{j}=(\nabla_1 e_+)_1~,~~~\delta^{ij} (\nabla_i e_1)_{j}=-{1\over2}[(\nabla_- e_-)_1+
(\nabla_+ e_+)_1]~,~~~
\cr
&&X_1=2[(\nabla_+ e_1)_+-(\nabla_- e_1)_-]~,~~~\tilde d \star\varphi=0~.
\la{g2gcona}
\eea
Observe that the above constraints imply that  $f$,
$e^+$, $e^-$, $e^1$ and $\varphi$ are all invariant under the action of $V$. Therefore $V$ preserves the $G_2$ structure of spacetime.

One can introduce some special coordinates on the spacetime. Solving the closure conditions, we write
 $e^1=f^{-2} dx^2$ and $e^5= f^{-2} dx^1$, and adapt a coordinate $t$ along the Killing vector field $V$,
 $V=\partial/\partial t$, where the functions $x^1$ and $x^2$ can be thought of as
spacetime coordinates. The metric of the spacetime $M$ can be written as
\bea
ds^2=-f^4 (dt+ m)^2+ f^{-4} \sum_{S=1}^2(dx^S)^2+ ds^2_{7}~,~~~ds^2_{7}=\delta_{ij} e^i e^j~,
\la{g2metr}
\eea
where $ds^2_{7}$ is the metric along the  ``transverse'' directions, $e^i=e^i_S dx^S+ e^i_I dy^I$,  $m=V_S dx^S+m_I dy^I$, and $y^I$
are the remaining coordinates of the spacetime.

The spacetime is foliated by  eight-dimensional Lorentzian manifolds $N$ given by $x^S={\rm const}$.
In turn $N$
is a real line bundle over a seven-dimensional base space $B$. A special case arises whenever $m_I=0$. In such a case $M$
is foliated by seven-dimensional manifolds $B$ given by
 $t, x^S={\rm const}$.  Moreover the conditions (\ref{g2gcona}) imply that $B$ is a co-symplectic
 $G_2$ manifold. This is perhaps the most significant geometric property of this class
 of IIB supersymmetric backgrounds. As we shall see the D3-brane and the LLM solutions are of this type.


\subsection{ Fluxes}

To find the conditions that supersymmetry imposes on the flux $F$ in (\ref{mfg2}), it is convenient to write
\bea
\label{fluxdecompg2}
\Phi= e^1\wedge \a-\star \a~,~~~\Psi=e^1\wedge \b+\star\b~,~~~{\cal X}=e^1\wedge \g+\d~,
\eea
where $\a, \b, \d$ are three-forms and $\g$ is a two-form.
Then using the decomposition $\Lambda^3(\bR^7)=\Lambda^3_{{\bf 1}}\oplus \Lambda^3_{{\bf 7}}\oplus \Lambda^3_{{\bf 27}}$ in $G_2$ irreducible
representations,
   $\a, \b, \d$ can be written as
\bea
&&\a=p_1\, \varphi+ v_1\bar\wedge \star \varphi+s_1\bar\wedge \varphi~,~~~\b=p_2\, \varphi
+ v_2\bar\wedge \star \varphi+s_2\bar\wedge \varphi
\cr
&&\d=p_3\, \varphi+ v_3\bar\wedge \star \varphi+s_3\bar\wedge \varphi~,
\eea
respectively,
where $s_1, s_2$ and $s_3$ are symmetric traceless 2-tensors associated with the  ${\bf 27}$-dimensional representation of $G_2$.
In particular, one has
\bea
p_1={1\over42} \varphi^{ijk} \a_{ijk}~,~~~(v_1)_i=-{1\over24} \a_{jkl} \star\varphi^{jkl}{}_i
\cr
(s_1)_{ij}={1\over4} \a_{kl(i} \varphi^{kl}{}_{j)}-{1\over28} \delta_{ij}\,\, \a_{klm} \varphi^{klm}
\eea
and similarly for the rest.
Using $\Lambda^2(\bR^7)=\Lambda_{{\bf 7}}\oplus \mathfrak{g}_2$, one has
\bea
\g=u\bar\wedge \varphi+ \hat\g~,
\eea
where $\hat\g\in \mathfrak{g}_2$ and
\bea
u_i={1\over6} \g_{jk} \varphi^{jk}{}_i~,~~~\hat\gamma_{ij}=\gamma^{\mathfrak {g}_2}_{ij}
=(\Pi_{{\bf 14}} \gamma)_{ij}~.
\eea

The IIB Killing spinor equations imply that all the five-form flux is determined in terms of the geometry. In particular
a calculation reveals that
\bea
p_1=\pm{\sqrt {2}\over 14} (\nabla_+ e_+)_1~,~~~(v_1)_i=\pm {\sqrt {2}\over 8} (\nabla_+ e_+)_i~,
\cr
(s_1)_{ij}=\mp {\sqrt {2}\over 8}\big[{1\over2}
(s_\Gamma)_{ij}-\big((\nabla_{(i} e_1)_{j)}-{\delta_{ij}\over7} (\nabla_k e_1)^k\big)\big]
\cr
p_2=\mp{\sqrt {2}\over 14} (\nabla_- e_-)_1~,~~~(v_2)_i=\pm {\sqrt {2}\over 8} (\nabla_- e_-)_i~,
\cr
(s_2)_{ij}=\mp {\sqrt {2}\over 8}\big[{1\over2} (s_\Gamma)_{ij}+\big((\nabla_{(i} e_1)_{j)}-{\delta_{ij}\over7}
(\nabla_k e_1)^k\big)\big]
\cr
p_3=\pm {\sqrt {2}\over 7} (\nabla_- e_-)_+~,~~~(v_3)_i=\mp {\sqrt {2} \over 8} (\nabla_+ e_1-\nabla_- e_1)_i
\cr
(s_3)_{ij}=\pm {\sqrt {2}\over 4} \big[(\nabla_{(i} e_+)_{j)}-{\delta_{ij}\over7} (\nabla_k e_+)^k\big]~,
\eea
 where $s_\Gamma$ is defined in the appendix \ref{g2}.
Similarly, we get
\bea
u_i=\pm{\sqrt{2}\over 12} \big((\nabla_j e_+)_k-(\nabla_je_-)_k\big) \varphi^{jk}{}_i
\cr
\hat\gamma_{ij}=\mp {1\over 2 \sqrt{2}} [(\nabla_{[i} e_+)_{j]}-(\nabla_{[i} e_-)_{j]}]^{\mathfrak {g}_2}~.
\eea

Substituting these expressions back into the flux and after some computation, one finds that
\bea
F = \Theta+ {}^*\Theta~,
\eea

where
\bea
\Theta =\mp {1\over4} \big{\{} f^{-4} \kappa
\wedge d(f^2 \varphi)+f^{-2} e^+ \wedge e^- \wedge e^1 \wedge d \kappa
 \big{\}}~,
 \la{f5g2}
\eea
and $\kappa = f^2 e^0$.
This is a remarkably simple expression for the flux. To summarize the solution
of the Killing spinor equations is given by (\ref{g2metr}) and (\ref{f5g2}) subject to the
conditions (\ref{g2gcona}).

\newsection{Co-calibrated $G_2$ manifolds and supergraviy backgrounds}

\subsection{Co-calibrated $G_2$ manifolds}

Co-calibrated or co-symplectic $G_2$ manifolds are seven-dimensional manifolds with a $G_2$ structure
which have the property that
\bea
\tilde d\star\varphi=0~,
\eea
i.e.~the dual of the fundamental $G_2$  form is closed. If one in addition requires that $X_3=0$, then from the results
of appendix  \ref{g2} it follows that $\tilde d \varphi=\l \, \star\varphi$,
$\l\not=0$. These are the {\it weak holonomy}  or {\it nearly parallel} $G_2$ manifolds. So the co-symplectic $G_2$
manifolds include the nearly parallel $G_2$ manifolds, which in turn include the tri-Sasakian ones.

Nearly parallel $G_2$ manifolds have appeared before
 in the context of eleven-dimensional supergravity compactifications, see e.g.~\cite{romans, romansw}. Compact
 homogeneous examples are the squashed 7-sphere, $Sp(2)\times Sp(1)/Sp(1)\times Sp(1)$, the Aloff-Wallach spaces $N(k,\ell)=SU(3)/U(1)$ \cite{aloff}
 and $SO(5)/SO(3)$ \cite{romansw}, where the embedding of $U(1)$ in $SU(3)$ is given as ${\rm diag}(e^{ik\theta_1}, e^{i\ell\theta_2}, e^{-i(\theta_1+\theta_2)})$
 and  $SO(3)$ is maximal in $SO(5)$. These are the only strictly nearly parallel $G_2$ compact homogeneous
 manifolds \cite{nps}. However, there are additional homogeneous
  tri-Sasakian and Einstein-Sasakian manifolds, see e.g.~\cite{duffpope, castellani, nps}. More examples of nearly parallel $G_2$ manifolds can be constructed
  by squashing the tri-Sasakian ones along the orbits of $\mathfrak{so}(3)$\cite{gsalamon, nps}. Using the tri-Sasakian examples of \cite{boyer},
  one can construct infinite families of such manifolds which are neither  homogeneous nor tri-Sasakian.

There are co-symplectic $G_2$ manifolds which are {\it not} nearly parallel. This can be easily seen from the results of \cite{swann}.
Unlike the nearly parallel case, there is no classification of compact homogeneous co-symplectic $G_2$ manifolds. It is expected that
the class of compact homogeneous co-symplectic $G_2$ manifolds is
large\footnote{We thank R. Cleyton, S. Ivanov and A. Swann  for discussions on this point.}.

The co-homogeneity one
co-symplectic $G_2$ manifolds are more tractable. A classification has been given in \cite{swann}. It has been found that
the principal orbits, which are six-dimensional homogeneous manifolds, are one of the following spaces
\bea
S^6=G_2/SU(3)~,~~~\bC P^3=Sp(2)/SU(2) U(1)~,~~~F_{1,2}=SU(3)/T^2~,
\cr
S^3\times S^3=SU(2)^3/SU(2)=SU(2)^2 T^1/T^1=SU(2)^2~,
\cr
S^5\times S^1=SU(3) T^1/SU(2)~,~~~S^3\times (S^1)^3=SU(2) T^3~,~~~(S^1)^6=T^6~,
\la{prin}
\eea
up to discrete identifications.
The metric and fundamental form of co-homogeneity one    manifolds can be written as
\bea
ds_{7}^2= h^2(y)dy^2+ ds^2_6(\mu(y))~,~~~\varphi=\varphi(h(y), \mu(y), \theta(y))~,
\eea
where $\mu$ are the homogeneous moduli of the $G_2$ structure which have been promoted to functions of the inhomogeneous  coordinate $y$.
We shall explain the $\theta$ dependence later.
Using a $y$ coordinate transformation, one can set $h=1$ as was done in \cite{swann}. However, we shall not do so
because in the associated supergravity backgrounds $h$ may depend on another two coordinates. Setting $e^6=h dy$, it is known
that the stability subgroup of a vector in the seven-dimensional representation of $G_2$ is $SU(3)$. There
is a unique fundamental $SU(3)$ two-form and the space of fundamental $SU(3)$  3-forms is two-dimensional. Using this,
the $G_2$ invariant three form can be written as
\bea
\varphi={\rm Re}(e^{i\theta} \hat\chi)+e^6\wedge \hat\omega~,
\la{cofun}
\eea
where $\hat\omega$ is the Hermitian form and $\hat\chi$ is the (3,0)-form.
The isotropy groups of the homogeneous spaces may leave more forms invariant, and so there may be many ways
to construct a $G_2$ invariant 3-form $\varphi$. As we shall demonstrate below, the bubbling $AdS$ solutions are examples
of backgrounds based on co-symplectic co-homogeneity one $G_2$ manifolds whose principal orbit is $S^3\times S^3$.
In fact one uses a very special family of such manifolds with isometry $SO(4)\times SO(4)$. Solutions
based  on various families with principal orbits given in (\ref{prin}) will be presented elsewhere \cite{ujg}.

\subsection{$D3$ brane}

The D3-brane solution can be viewed as a special case of solutions with $G_2$-invariant Killing spinors.
For this write the metric as
\bea
ds^2= -h^{-{1\over2}} dt^2+ h^{{1\over2}} (dx^2+dy^2)+ h^{{1\over2}} dw^2+ h^{-{1\over2}} d{\bf z}^2+h^{{1\over2}} d{\bf q}^2~,
\eea
where $(t, {\bf z})$ are the worldvolume and $(x,y,w, {\bf q})$ are the transverse coordinates, respectively
and $h$ is the usual harmonic function $h=h(w,x,y, {\bf q})$. The time-like Killing vector field is $\kappa=\partial/\partial t$ and
so $f^4= h^{-{1\over2}}$.  The fundamental $G_2$ three-form is
\bea
\varphi={\rm Re} \prod^3_{r=1} [\wedge (h^{-{1\over4}} d{\bf z}^r+i h^{{1\over4}} d{\bf q}^r)]-
h^{{1\over4}} dw \wedge \sum_{r=1}^3 d{\bf z}^r\wedge d{\bf q}^r~,
\eea
and $e^1= h^{{1\over4}} dx$ and $e^5=h^{{1\over4}} dy$. It is straightforward to verify that
$\tilde d\star\varphi=0$. Moreover $Y_r=X_1=0$. The remaining conditions are also satisfied. A consequence of this
is that the seven-dimensional ``transverse'' space $B$ admits a  $G_2$ structure for which the only non-vanishing
class is $X_3$. Asymptotically, $B$ is flat space and at the origin is $H_4\times S^3$.

\subsection{Bubbling AdS Solutions}

The   solution found by Lin, Lunin and Maldacena in
\cite{llmsol} is an example of a $G_2$ IIB supergravity background and so it is associated with
a co-symplectic $G_2$ geometry. The co-symplectic geometry is of co-homogeneity one and has principal orbit $S^3\times S^3$.
To show this, we write the metric in the form given in (\ref{g2metr}) by introducing a frame
\bea
e^0 &=& h^{-1}(dt+V_S dx^S) ~,~~~e^5= h dx^1~,~~~e^1 = h dx^2~,~~~
\cr
e^6 &=& h dy~,~~~e^{1+a}={\sqrt{y} \over 2} e^{{G \over 2}} \sigma^a~,~~~e^{6+a}={\sqrt{y} \over 2} e^{-{G \over 2}} {\hat{\sigma}}^a~,
\eea
where $\sigma^a, {\hat{\sigma}}^a$, $a=1,2,3$, are  left-invariant 1-forms
on $SU(2)\times SU(2)$ given by
\bea
\sigma^1 &=& - \sin \psi d \theta + \cos \psi \sin \theta d \phi~,~~~
\sigma^2 = \cos \psi d \theta + \sin \psi \sin \theta d \phi~,
\cr
\sigma^3 &=& d \psi + \cos \theta d \phi~,
\eea
and ${\hat{\sigma}}^a$ are defined in exactly the same way, but with
$\theta, \phi, \psi$ replaced with ${\hat{\theta}}$, ${\hat{\phi}}$,
${\hat{\psi}}$ throughout, and the rest of the components of the metric depend on $y, x^S$. Moreover, the functions $h$ and $G$ are
related by
\be
h^{-2} = 2y \cosh G~,
\ee
and
\bea
\label{llmcon1}
&&{\partial V_1 \over \partial y} = {1 \over 2y \cosh^2 G} {\partial G \over \partial x^2}~,~~~
{\partial V_2 \over \partial y} =-{1 \over 2y \cosh^2 G} {\partial G \over \partial x^1}
\cr
&&{\partial V_2 \over \partial x^1} - {\partial V_1 \over \partial x^2}
={1 \over 2y \cosh^2 G} {\partial G \over \partial y}~.
\eea
Writing $z={1 \over 2} \tanh G$, $z$ is further constrained by
\be
\partial_S \partial_S z +y \partial_y \big( {\partial_y z \over y} \big) =0~.
\ee

The five-form is given by
\be
F = {1 \over 8} F_{(1)} \wedge \sigma^1 \wedge \sigma^2 \wedge \sigma^3
-{1 \over 8} F_{(2)} \wedge {\hat{\sigma}}^1 \wedge
{\hat{\sigma}}^2 \wedge {\hat{\sigma}}^3~,
\ee
where $F_{(1)}$, $F_{(2)}$ are two 2-forms. In turn these are given by
\be
F_{(1)} = d B_t \wedge (dt + V_S dx^S) + B_t dV + d {\tilde{B}}~,
\ee
with
\be
B_t = -{1 \over 4} y^2 e^{2G}, \qquad d {\tilde{B}} = -{1 \over 4} y^3
\star_3 d({z+{1 \over 2} \over y^2})~,
\ee
and
\be
F_{(2)} = d C_t \wedge (dt + V_S dx^S) + C_t dV +d {\tilde{C}}~,
\ee
with
\be
C_t = -{1 \over 4} y^2 e^{-2G}, \qquad d {\tilde{C}} = -{1 \over 4} y^3
\star_3 d({z-{1 \over 2} \over y^2})~,
\ee
where $\star_3$ denotes Hodge duality on $\bR^3$ with metric
$dy^2+(dx^1)^2+(dx^2)^2$ with positive orientation given by $dy \wedge dx^1 \wedge dx^2$.

In this basis the transverse  metric is
\be
ds^2_{7} = (e^2)+(e^3)^2+(e^4)^2+(e^6)^2+(e^7)^2+(e^8)^2+(e^9)^2~,
\ee
equipped with the fundamental $G_2$ form
\be
\varphi = {\rm Re \ } \big( e^{iH} (e^2+ie^7) \wedge (e^3+ie^8) \wedge (e^4+ie^9) \big)
- e^6 \wedge (e^2 \wedge e^7 + e^3 \wedge e^8 + e^4 \wedge e^9)~,
\ee
where
\be
e^{iH} = {1 \over \sqrt{e^G+e^{-G}}}(-e^{{G \over 2}}+i e^{-{G \over 2}})~.
\ee
Observe that the expression for the fundamental form is as expected in the context
of co-homogeneity one $G_2$ structures (\ref{cofun}).
The frame we have chosen is not adapted in the $G_2$ structure because the components of
$\varphi$ depend on $H$. However, we can adapt
a frame by rotating the frame as
\bea
e^2+ie^7 \rightarrow  e^{{iH \over 3}}(e^2+ie^7)~,
\eea
and similarly for the other two pairs while leaving $e^6$ as it is. Observe that the metric remains invariant under this rotation.
However, in what follows in this section we shall continue  to work with the frame we have originally introduced.

It is straightforward to verify that $\varphi$ satisfies
\be
{\tilde{d}} \star_7 \varphi =0~.
\ee
This establishes that the LLM solution is associated with a co-symplectic $G_2$ structure. Moreover, it is straightforward
to see by setting $y, x^S={\rm const}$ that the principal orbit is $S^3\times S^3$. Observe that this principal orbit is one of the
possibilities that appear in the classification of co-homogeneity one co-symplectic manifolds in \cite{swann} and are
listed in (\ref{prin}). In fact, it is associated with the most symmetric family
of such co-symplectic structures which has $SO(4)\times SO(4)$ symmetry. There are families of the same orbit that possess fewer isometries.
Although $\varphi$ is co-closed on the transverse space, ${\tilde{d}} \varphi \neq 0$,
and we find that
\be
X_1 = {4 \sqrt{y}  \over \sqrt{e^G+e^{-G}}} {\partial G \over \partial y}~.
\ee
The function $f$ appearing in (\ref{g2metr}) is given by
\be
f^4 = 2 y \cosh G~.
\ee

It remains to find the forms $\alpha, \beta, \delta$ which appear   in the
decomposition of $F$,  ({\ref{fluxdecompg2}}). These are
\bea
\alpha &=&{1 \over 2 \sqrt{2y}}e^{-{G \over 2}} \big(ye^G{\partial G \over \partial x^2}
+1-y {\partial G \over \partial y} \big) e^2 \wedge e^3 \wedge e^4
\cr
&+&{1 \over 2 \sqrt{2y}} e^{{G \over 2}} \big(ye^{-G}{\partial G \over \partial x^2}
+1+y{\partial G \over \partial y} \big) e^7 \wedge e^8 \wedge e^9~,
\cr
\beta &=&{1 \over 2 \sqrt{2y}}e^{-{G \over 2}} \big(-ye^G{\partial G \over \partial x^2}
+1-y {\partial G \over \partial y} \big) e^2 \wedge e^3 \wedge e^4
\cr
&+&{1 \over 2 \sqrt{2y}} e^{{G \over 2}} \big(-ye^{-G}{\partial G \over \partial x^2}
+1+y{\partial G \over \partial y} \big) e^7 \wedge e^8 \wedge e^9~,
\eea
and
\be
\delta = {1 \over 2} \sqrt{y} e^{{G \over 2}} {\partial G \over \partial x^1}
e^2 \wedge e^3 \wedge x^4 + {1 \over 2} \sqrt{y} e^{-{G \over 2}} {\partial G \over \partial x^1}
e^7 \wedge e^8 \wedge x^9~.
\ee
Moreover,  $\gamma$ vanishes.
It is then straightforward  to verify, using
({\ref{llmcon1}}), that all of the remaining geometric
constraints are satisfied, and the
components of $F$ are determined in terms of the spin connection
as set out in the previous section.

There are other solutions which are extensions of the LLM solution and so they are associated with
a co-symplectic $G_2$ structure. In particular observe that  (\ref{g2metr}) is compatible with some
recent results in \cite{chen} where bubbling solutions
have been investigated that preserve 4, 8 and 16 supersymmetries.

\newsection{Concluding remarks}

IIB backgrounds with $N\geq 2$ supersymmetry and active five-form flux have four distinct types of geometries. Killing spinors
with isotropy group $SU(4)\ltimes\bR^8$ give rise to two types of geometry depending on whether the Killing spinors
are generic or pure. In both cases, the spacetime admits a null Killing vector field and an almost complex structure on
the transverse directions to the lightcone. For generic backgrounds, the twist of  the
Killing vector field is trivial but the almost complex structure non-integrable, while in the pure case the twist is non-trivial
but the almost complex structure is integrable. In the latter case, if one assumes that the twist of the Killing vector
field is trivial, then all the geometric conditions take a very simple form. In particular, the cubic power of a certain rescaled
Hermitian form and a certain rescaled fundamental (4,0)-form must be closed. Examples of backgrounds that admit pure
Killing spinors are the D3-brane and the lifts of supersymmetric backgrounds of gauged five-dimensional supergravity which include the
$AdS_5$ black holes. The null Killing vector field of the D3-brane background has a trivial twist but that of the $AdS_5$ black hole does not.
$N\geq 2$ backgrounds with Killing spinors
which have isotropy group $Spin(7)\ltimes \bR^8$ are pp-waves propagating on a manifold with holonomy contained in $Spin(7)$.

The remaining type of geometry is associated with IIB backgrounds that admit $G_2$-invariant Killing spinors. The tangent
bundle of such spacetimes decomposes with respect to the spacetime metric into a trivial bundle of rank three and a bundle
of rank seven which are the transverse directions of the spacetime. One direction in the trivial bundle is associated
with a time-like Killing vector field and the other two directions are associated with closed one forms. The transverse geometry is that
of a co-symplectic $G_2$ manifold. We give a  description of how one can construct
a supersymmetric IIB background from a family of co-symplectic $G_2$ manifolds .  In addition, we
 show that the co-symplectic $G_2$ manifold associated with  the bubbling $AdS$ solution of \cite{llmsol}
has principal orbit  $S^3\times S^3$. The transverse structures of the supersymmetric backgrounds are summarized in table 1.

\begin{table}[ht]
 \begin{center}
\begin{tabular}{|c|c|c|}\hline
 ${\mathrm Stab}$ & {\rm Transverse structure}&{\rm Conditions}
 \\ \hline \hline  {\rule{0pt}{3.0ex}}
 $ SU(4)\ltimes\bR^8$&{\rm Relatively balanced }& $\tilde d\omega^3=W_4\wedge \omega^3$~,~~$W_1=W_2=0$
 \\
 {\rm pure}&{\rm $SU(4)$  Hermitian}& $W_4=W_5$
\\ \hline {\rule{0pt}{3.0ex}}
 $Spin(7)\ltimes\bR^8$& {\rm Holonomy}~ $Spin(7)$& $\tilde d\phi=0$\\
\hline {\rule{0pt}{3.0ex}}
$G_2$&{\rm co-symplectic} $G_2$& $\tilde d\star\varphi=0$
\\ \hline
\end{tabular}
\end{center}
\caption{The  columns contain the stability subgroups of the Killing spinors in $Spin(9,1)$, the type of transverse geometry, and
the conditions on the fundamental forms, respectively.}
\end{table}

The holonomy of the supercovariant connection of IIB supergravity with only $F$ flux is contained in $SL(16, \bC)$.
This can be easily seen
from the computation of the super-covariant curvature for this case in \cite{jfgpa}. This is   a subgroup of
$SL(32,\bR)$ which is the holonomy group of the IIB supercovariant connection \cite{tsimpis}. Arguments based on the reduction
of holonomy in the presence of parallel spinors similar to those of  \cite{hull, duff} suggest that there may exist supersymmetric
IIB  backgrounds with $F$ flux
with any even number of supersymmetries.  We have identified the geometries of $N=2$ backgrounds and the
$N=32$, maximally supersymmetric, backgrounds have been classified in \cite{jfgpa}.
It is also tractable to investigate the existence of $N=30$ backgrounds by applying  the
 technique developed in \cite{preons} to this case. If  backgrounds exist with strictly $N=30$ supersymmetry,
 they are expected to be homogeneous spaces \cite{jfof}. We hope to report the outcome of this investigation in the future.

\vskip 0.5cm {\bf Acknowledgments:} U.G.~is supported by the Swedish Research Council. We thank Diederik
Roest for many useful discussions.

\setcounter{section}{0}

\appendix{Linear Systems}
\setcounter{subsection}{0}

\subsection{Solution of the linear system of generic $SU(4)\ltimes\bR^8$ backgrounds}
\la{lsgsu4}

The Killing spinor is given in (\ref{kssu4}). For generic backgrounds, we take $f, g_1, g_2\not=0$ and $f\not=\pm g_2$.
The conditions that arise from the Killing spinor equations for supersymmetric backgrounds with $F$ flux,
$P=G=0$, and $SU(4)\ltimes\bR^8$-invariant Killing spinors can be easily read from those of $N=1$ backgrounds
in \cite{ugjggpa}.  First consider the conditions that arise from the gravitino Killing spinor equation ${\cal D}_+\epsilon=0$.
These can be written as
\bea
&&\Omega_{+,\a\b}=0~,~~~\Omega_{+,\b}{}^\b=0~,~~~\partial_+f+{1\over2}\Omega_{+,-+} f=0~,
\cr
&&\partial_+g_1+{1\over2}\Omega_{+,-+} g_1=0~,~~~\partial_+g_2+{1\over2}\Omega_{+,-+} g_2=0~,
\cr
&&\Omega_{+,+\a}=\Omega_{+,+\bar\a}=0~.
\eea

Some of the conditions associated with the gravitino Killing spinor equations ${\cal D}_\a\epsilon={\cal D}_{\bar\a}\epsilon=0$
give
\bea
&&\Omega_{\a,+\b}=0~,~~~F_{+\bar\a\b_1\b_2\b_3}=0~,
~~~\Omega_{\bar\a,+\b}=0~,~~~F_{+\bar\a\b\g}{}^\g=0~,
\cr
&&
2 \Omega_{\bar\a, \b_1\b_2} (f^2-(g_2+ig_1)^2)-(f^2+g_1^2+g_2^2) \Omega_{\bar\a, \bar\g_1\bar\g_2} \epsilon^{\bar\g_1\bar\g_2}{}_{\b_1\b_2}=0~,
\cr
&&
4i F_{-+\bar\a\b_1\b_2}+4i g_{\bar\a[\b_1} F_{\b_2]-+\d}{}^\d+ 2 \Omega_{\bar\a, \bar\g_1\bar\g_2}
\epsilon^{\bar\g_1\bar\g_2}{}_{\b_1\b_2} {f g_2\over f^2-(g_2+i g_1)^2}=0~.
\la{omcon}
\eea



Next some of the conditions of ${\cal D}_-\e=0$, their complex conjugate and dual give
\bea
&&-4i f g_2 F_{-\bar\a_1\bar\a_2\b}{}^\b+ 2 [f^2+g_1^2+g_2^2] \Omega_{-,\bar\a_1\bar\a_2}
-[f^2-(g_2+i g_1)^2] \Omega_{-, \g_1\g_2} \epsilon^{\g_1\g_2}{}_{\bar\a_1\bar\a_2}=0~,
\cr
&& [f^2+g_1^2+g_2^2]\Omega_{-,+\bar\a}-2i f g_2 F_{-+\bar\a\b}{}^\b=0~,
 \cr
 &&
 [f^2-(g_2-i g_1)^2] \Omega_{-,+\bar\a}-{2i\over3} f g_2 F_{-+\g_1\g_2\g_3} \epsilon^{\g_1\g_2\g_3}{}_{\bar\a}=0~.
 \la{omegaf}
 \eea
Some of the ${\cal D}_-\e=0$ and some of the remaining ${\cal D}_\a\e={\cal D}_{\bar\a}\e=0$ conditions give
\bea
[D_{\bar\a}+{1\over2} \Omega_{\bar\a,\b}{}^\b+{1\over2} \Omega_{\bar\a,-+}+\Omega_{-,+\bar\a}+i F_{-+\bar\a\b}{}^\b](f-g_2+ig_1)&=&0~,
\cr
[D_{\bar\a}+{1\over2} \Omega_{\bar\a,\b}{}^\b+{1\over2} \Omega_{\bar\a,-+}+\Omega_{-,+\bar\a}-i F_{-+\bar\a\b}{}^\b](f+g_2-ig_1)&=&0~,
\cr
D_{\bar\a} (f^2+g_1^2+g_2^2)+[\Omega_{\bar\a,-+}+\Omega_{-,\bar\a+}] (f^2+g_1^2+g_2^2)&=&0~.
\la{1ba3a}
\eea
In addition, one finds that
\bea
D_{\bar\a}\log[f^2-(g_2+i g_1)^2]-\Omega_{\bar\a,\b}{}^\b+\Omega_{\bar\a,-+}&=&0~,
\cr
D_{\bar\a}\log[{f+g_2+i g_1\over f-g_2-i g_1}]-2i F_{-+\bar\a\b}{}^\b&=&0~.
\eea
{}From (\ref{1ba3a}), one gets
\bea
D_{\bar\a}\log[f^2-(g_2-i g_1)^2]+\Omega_{\bar\a,\b}{}^\b+\Omega_{\bar\a,-+}+2\Omega_{-,+\bar\a}&=&0~,
\cr
D_{\bar\a}\log[{f-g_2+i g_1\over f+g_2-i g_1}]+2i F_{-+\bar\a\b}{}^\b&=&0~.
\eea
Thus
\bea
D_{\bar\a}\log[{(f+i g_1)^2-g^2_2\over (f-i g_1)^2-g^2_2}]=0~,
~~~4i F_{-+\bar\a\b}{}^\b+D_{\bar\a}\log{(f-g_2)^2+g_1^2\over (f+g_2)^2+g_1^2}=0~.
\eea
Comparing  with (\ref{omegaf}), one finds that
\bea
\Omega_{-,+\bar\a}-{ f g_2\over 2(f^2+g_1^2+g_2^2)} D_{\bar\a}\log[{(f+g_2)^2+ g_1^2\over (f-g_2)^2+ g_1^2}]=0~.
\eea
Furthermore
\bea
D_{\bar\a}\log[((f-g_2)^2+g_1^2) ((f+g_2)^2+g_1^2)]+2 \Omega_{\bar\a,-+}+ 2\Omega_{-,+\bar\a}=0~,
\cr
D_{\bar\a}\log[{f^2-(g_2-i g_1)^2\over f^2-(g_2+i g_1)^2}]+2\Omega_{\bar\a,\b}{}^\b+ 2\Omega_{-,+\bar\a}=0~.
\eea
Taking the trace of the last two equations in (\ref{omcon}),  one gets the additional geometric conditions
\bea
-\Omega_{\bar\b,\a}{}^{\bar\b}+{(f^2+g_1^2+g_2^2)^2\over 4 f^2 g_2^2} \Omega_{-,+\a}=0~,
\cr
{( f^2+g_1^2+g_2^2)\over 2f g_2}\Omega_{-,+\a}+
{f g_2\over f^2-(g_2+i g_1)^2}\, \Omega_{\bar\g_1,\bar\g_2\bar\g_3} \epsilon^{\bar\g_1\bar\g_2\bar\g_3}{}_\a=0~.
\eea
The remaining ${\cal D}_-\e=0$ equations give
\bea
&&\partial_-(f^2-(g_2-i g_1)^2)+{i\over3} f g_2 F_{-\g_1\dots\g_4} \epsilon^{\g_1\dots \g_4}+ [\Omega_{-,\g}{}^\g+\Omega_{-,-+}]
(f^2-(g_2-i g_1)^2)=0~,
\cr
&&
-2i g_2 \partial_- g_1 +2i g_1 \partial_- g_2
+ (f^2+g_1^2+g_2^2) \Omega_{-,\alpha}{}^\alpha
-if g_2 F_{- \alpha}{}^\alpha{}_\beta{}^\beta=0~,
\cr
&&\partial_-[f^2+g_1^2+g_2^2]+\Omega_{-,-+} [f^2+g_1^2+g_2^2]=0~.
\eea
We have not been able to simplify the solution of the linear system further. These conditions have been expressed
in terms of the fundamental $SU(4)$ forms in section \ref{sgsu4}.

\subsection{Pure $SU(4)\ltimes\bR^8$  Killing spinors}
\subsubsection{Solution of the linear system }
\la{lspsu4}

We choose the Killing spinor to be $\e=h 1$. Then after  some straightforward computation using the
results of \cite{ugjggpa}, the linear system arising from the Killing spinor equations can be solved.
The conditions on the geometry are
\bea
&&\partial_+h+{1\over2} \Omega_{+,-+} h=0~,~~~\partial_-h+{1\over2} \Omega_{-,-+}h=0~,~~~
\partial_\a h+{1\over2} (\Omega_{\a,-+}+\Omega_{-,\a+}) h=0~,~~~
\cr
&&\Omega_{+,\b}{}^\b=0~,~~~\Omega_{+,\a\b}=0~,~~~\Omega_{\a,+}{}^\a=0~,
\cr
&&\Omega_{+,+\a}=0~,~~~\Omega_{\a,+\b}=0~,~~~\Omega_{\a,+\bar\b}+\Omega_{\bar\b,+\a}=0~,
\cr
&&\Omega_{\a,\b\g}=0~,~~~~\Omega_{\bar\a,\b}{}^\b+\Omega_{\b,\bar\a}{}^\b=0~,~~~\Omega_{\bar\a,\b}{}^\b=-\Omega_{-,+\bar\a}~.
\eea
In addition, one finds the following restrictions on the fluxes
\bea
&&F_{+\a\bar\b_1\bar\b_2\bar\b_3}=0~,~~~i F_{\a+\bar\b\g}{}^\g+\Omega_{\a,+\bar\b}=0~,~F_{\a\bar\b_1\bar\b_2\bar\b_3\bar\b_4}=F_{-+\bar\b_1\bar\b_2\bar\b_3}=0~,
\cr
 &&iF_{\bar\a-+\b}{}^\b=\Omega_{\bar\a,\b}{}^\b~,
~~~
2iF_{\a-+\bar\b_1\bar\b_2}-2i g_{\a[\bar\b_1} F_{\bar\b_2]-+\g}{}^\g+\Omega_{\a,\bar\b_1\bar\b_2}=0~,~~~F_{-\a_1\a_2\a_3\a_4}=0~,
\cr
&&\Omega_{-,\bar\b\bar\g}+i F_{-\bar\b\bar\g\d}{}^\d=0~,~~~
\Omega_{-,\b}{}^\b+{i\over2}F_{-\a}{}^\a{}_\b{}^\b=0~,~~~
\Omega_{-,+\a}+i F_{-+\a\b}{}^\b=0~.
\eea
 The above conditions on the geometry  and the fluxes imposed by supersymmetry have been re-expressed in terms of the
 fundamental $SU(4)$ forms in section \ref{spsu4}.

\subsubsection{Spin connection}
\la{spincon}

The non-vanishing components of the spin connection
along the transverse directions of the spacetime associated with the uplift of the
five-dimensional supersymmetric solution can be expressed as
\bea
 \Omega_{a,\nu {\bar{\rho}}} &=& -{2i \over 3} k_2 {\cal{P}}_a \delta_{\nu {\bar{\rho}}}~,~~~
 \Omega_{\mu,b {\bar{c}}} = -{2i \over 3} k_1 {\cal{Q}}_\mu \delta_{b \bar{c}}~,
\cr
 \Omega_{a,bc} &=& e^{-3ik_1 \chi_2} \cF^{1 \over 2}
 {\hat{\Omega}}_{\hat{a}, \hat{b} \hat{c}}~,
 \cr
 \Omega_{a,b \bar{c}} &=& -{1 \over 2} \cF^{-1} \nabla_b \cF
 \delta_{a \bar{c}} +{1 \over 2} \cF^{-1} \nabla_a \cF \delta_{b \bar{c}}
 -{2i \over 3} k_1 {\cal{P}}_a \delta_{b \bar{c}}
 +\cF^{1 \over 2} e^{-ik_1 \chi_2}  {\hat{\Omega}}_{\hat{a}, \hat{b}
 \bar{\hat{c}}}~,
 \cr
 \Omega_{a, \bar{b} \bar{c}} &=& {1 \over 2} \cF^{-1} \nabla_{\bar{b}}
 \cF \delta_{a \bar{c}} -{1 \over 2} \cF^{-1} \nabla_{\bar{c}} \cF
 \delta_{a \bar{b}}   + e^{ik_1 \chi_2} \cF^{1 \over 2}
 {\hat{\Omega}}_{\hat{a}, \bar{\hat{b}}  \bar{\hat{c}}}~.
 \eea
Here hatted and unhatted frame indices are with respect to the bases
 $e^a$ and ${\hat{e}}^a$, and ${\hat{\Omega}}_{\hat{a}, \hat{b} \hat{c}}$,
  ${\hat{\Omega}}_{\hat{a}, \hat{b}
 \bar{\hat{c}}}$, ${\hat{\Omega}}_{\hat{a}, \bar{\hat{b}}  \bar{\hat{c}}}$
 are the components of the spin connection of $N$ with respect to the basis
 ${\hat{e}}^a$. As $N$ is K\"ahler with K\"ahler form $J_N$, we have
 $ {\hat{\Omega}}_{\hat{a}, \hat{b} \hat{c}}=0$ and  ${\hat{\Omega}}_{\hat{a}, \bar{\hat{b}}  \bar{\hat{c}}}=0$.

 Similarly, we have
\bea
 \Omega_{\mu,\nu\rho} &=& e^{-3ik_2 \chi_2} \ell^{-1} {\hat{\Omega}}_{\hat{\mu}, \hat{\nu} \hat{\rho}}
 \cr
 \Omega_{\mu,\nu \bar{\rho}} &=& -{2i \over 3} {\cal{Q}}_\mu
 \delta_{\nu \bar{\rho}}+ e^{-ik_2 \chi_2}  \ell^{-1}
 {\hat{\Omega}}_{\hat{\mu}, \hat{\nu} \bar{\hat{\rho}}}
 \cr
 \Omega_{\mu, \bar{\nu} \bar{\rho}} &=& e^{ik_2 \chi_2} \ell^{-1}
 {\hat{\Omega}}_{\hat{\mu}, \bar{\hat{\nu}}  \bar{\hat{\rho}}}~,
 \eea
for the directions along $CP^2$.

\subsection{Solution of the linear system of $Spin(7)\ltimes\bR^8$ backgrounds}
\la{lsspin7}

The Killing spinor is $\e=(f+ig) (1+e_{1234})$, i.e.~it is given as in  (\ref{kssu4}) by setting $g_2=0$ and $g_1=g$.
A straightforward computation using the results of \cite{ugjggpa} reveals that the conditions on the geometry are
\bea
&&\partial_+(f+ig)+{1\over2} \Omega_{+,-+}(f+ig)=0~,~~~\partial_-(f^2+g^2)+\Omega_{-,-+} (f^2+g^2)=0~,~~~
\cr
&&\partial_{\bar\a} f+{1\over2}\Omega_{\bar\a,-+} f=0~,~~~\partial_{\bar\a} g+{1\over2}\Omega_{\bar\a,-+} g=0~,
\cr
&&\Omega_{+,\bar\b_1\bar\b_2}={1\over2} \epsilon_{\bar\b_1\bar\b_2}{}^{\g_1\g_2} \Omega_{+,\g_1\g_2}~,~~~\Omega_{+,\g}{}^\g=0~,
~~~
\Omega_{+,+\a}=0~,~~~\Omega_{\a,+\b}=0~,~~~\Omega_{\a,+\bar\b}=0~,
\cr
&&\Omega_{-,\b_1\b_2}-
{1\over2} \epsilon_{\b_1\b_2}{}^{\bar\g_1\bar\g_2} \Omega_{-, \bar\g_1\bar\g_2}=0~,~~~\Omega_{-,\b}{}^\b=0~,
~~~\Omega_{-,+\a}=0~,
\cr
&&\Omega_{\bar\a,\b_1\b_2}-{1\over2} \epsilon_{\b_1\b_2}{}^{\bar\g_1\bar\g_2} \Omega_{\bar\a, \bar\g_1\bar\g_2}=0~,~~~\Omega_{\bar\a,\b}{}^\b=0~,
\eea
and
 the conditions on the fluxes are
 \bea
 &&F_{+\bar\a\b_1\b_2\b_3}=0~,~~~F_{+\bar\a\b\g}{}^\g=0~,~~~ F_{-+\g_1\g_2\g_3} =0~,~~~ F_{-\bar\b_1\bar\b_2\d}{}^\d=-{1\over2} \epsilon_{\bar\b_1\bar\b_2}{}^{\g_1\g_2}
 F_{-\g_1\g_2\d}{}^\d~,
 \cr
&&F_{-+\bar\a\b_1\b_2}=0~,~~~
 \partial_-\log({f+ig\over f-ig})+{i\over2} F_{-\g}{}^\g{}_\d{}^\d+{i\over6} F_{-\g_1\dots\g_4} \epsilon^{\g_1\dots\g_4}~.
 \eea
As a special case suppose that a background admits a Majorana-Weyl Killing spinor. One can use the
gauge symmetry to write $\epsilon=f (1+e_{1234})$. The conditions are then easily derived
and they can be read from the ones above by setting $g=0$. The conditions above on the geometry and fluxes can be re-expressed in terms
of the fundamental $Spin(7)$ form in section \ref{sspin7}.

\subsection{Solution of the linear system of $G_2$ backgrounds}
\la{lsg2}

The Killing spinor is  $\epsilon= f(1+e_{1234})+ ig (e_{51}+e_{5234})$, where
$f$ and $g$ are real spacetime functions.
The equations that arise from the linear system are simplified considerably by making a gauge transformation of the form
$e^{b \Gamma_{+-}}$ to set $g^2=f^2$. We therefore take $g=\pm f$. Unlike the previous cases, the computation
for the $G_2$ case is more involved and it is convenient to organize the conditions in terms of $SU(3)\subset G_2$
representations. For this, we choose a pseudo-Hermitian frame $(e^+, e^-, e^1, e^{\bar 1}, e^p, e^{\bar p})$, $p=2,3,4$,
on the spacetime
which is adapted to the choice of the Killing spinors. The linear system can be easily derived
from that of \cite{ugjggpb} by setting the one-form and three-form field strengths to zero. The conditions that arise
from the linear system are as follows:

The conditions on the geometry and fluxes which transform as   $SU(3)$ singlets are
\bea
F_{+-1p}{}^p &=& \mp {1 \over 2} \big( \Omega_{1,-1} + \Omega_{\bar{1},-\bar{1}} +\Omega_{+,p}{}^p-\Omega_{-,p}{}^p \big)~,
\cr
F_{-1 \bar{1}p}{}^p &=& \pm {1 \over 4} \big( \Omega_{1,p}{}^p -\Omega_{\bar{1},p}{}^p +\Omega_{1,1\bar{1}}-\Omega_{\bar{1},1\bar{1}}
+\Omega_{-,-1}+\Omega_{-,-\bar{1}} \big)~,
\cr
F_{+1 \bar{1}p}{}^p &=& \pm {1 \over 4} \big(  \Omega_{1,p}{}^p -\Omega_{\bar{1},p}{}^p -\Omega_{1,1\bar{1}}+\Omega_{\bar{1},1\bar{1}}
 -\Omega_{+,+1}-\Omega_{+,+\bar{1}} \big)~,
\cr
F_{-1234} &=& \pm {1 \over 8} \big( -\Omega_{1,p}{}^p + \Omega_{\bar{1},p}{}^p - \Omega_{1,1\bar{1}}+\Omega_{\bar{1},1\bar{1}}
+3 \Omega_{-,-1}-\Omega_{-,-\bar{1}} \big)~,
\cr
F_{+-234} &=& \pm {1 \over 4} \big( \Omega_{1,-\bar{1}}+\Omega_{\bar{1},-1}+\Omega_{+,p}{}^p+\Omega_{-,p}{}^p \big)~,
\cr
F_{+1 \bar{2} \bar{3} \bar{4}} &=& \pm {1 \over 8} \big( -\Omega_{1,p}{}^p +\Omega_{\bar{1},p}{}^p +\Omega_{1,1\bar{1}}
-\Omega_{\bar{1},1\bar{1}} -3\Omega_{+,+1}+\Omega_{+,+\bar{1}} \big)~,
\cr
\Omega_{1,p}{}^p + \Omega_{\bar{1},p}{}^p &=& {1 \over 2} (\Omega_{+,+1}-\Omega_{+,+\bar{1}}-\Omega_{-,-1}
+\Omega_{-,-\bar{1}})~,
\cr
\Omega_{1,1\bar{1}}+\Omega_{\bar{1},1\bar{1}}&=&-{1 \over 2}(\Omega_{+,+1}-\Omega_{+,+\bar{1}}
+\Omega_{-,-1}-\Omega_{-,-\bar{1}})~,
\cr
\Omega_{1,-+}&=&{1 \over 2}( \Omega_{+,+1}- \Omega_{-,-1})~,
\cr
\Omega_{1,-\bar{1}}-\Omega_{\bar{1},-1}&=&- \Omega_{+,p}{}^p-\Omega_{-,p}{}^p~,
\cr
\Omega_{1,+1}&=&{1 \over 2} \big( \Omega_{1,-1} + \Omega_{\bar{1},-\bar{1}} +\Omega_{+,p}{}^p-\Omega_{-,p}{}^p \big)~,
\cr
\Omega_{1,+\bar{1}} &=& {1 \over 2} \big(\Omega_{1,-\bar{1}}+\Omega_{\bar{1},-1}+\Omega_{+,p}{}^p+\Omega_{-,p}{}^p \big)~,
\cr
\Omega_{p,}{}^p{}_{\bar{1}}+\Omega_{\bar{p},}{}^{\bar{p}}{}_1 &=&{1 \over 2} \big(\Omega_{1,p}{}^p
- \Omega_{\bar{1},p}{}^p +\Omega_{1,1\bar{1}}-\Omega_{\bar{1},1\bar{1}}~,
\cr
&-& \Omega_{p_1,p_2 p_3} \epsilon^{p_1 p_2 p_3} - \Omega_{\bar{p}_1,\bar{p}_2 \bar{p}_3} \epsilon^{\bar{p}_1 \bar{p}_2 \bar{p}_3}
+ \Omega_{-,-1}+\Omega_{-,-\bar{1}}\big)~,
\cr
\Omega_{p,}{}^p{}_+ &=& -{1 \over 2} \big( \Omega_{1,-1} + \Omega_{\bar{1},-\bar{1}} +\Omega_{+,p}{}^p-\Omega_{-,p}{}^p \big)~,
\cr
\Omega_{p,}{}^p{}_- &=& -{1 \over 2} \big( \Omega_{1,-1} + \Omega_{\bar{1},-\bar{1}} -\Omega_{+,p}{}^p+\Omega_{-,p}{}^p \big)~,
\cr
\Omega_{p_1,p_2 p_3} \epsilon^{p_1 p_2 p_3} - \Omega_{\bar{p}_1,\bar{p}_2 \bar{p}_3} \epsilon^{\bar{p}_1 \bar{p}_2 \bar{p}_3}&=&
-2(\Omega_{p,}{}^p{}_{\bar{1}}-\Omega_{\bar{p},}{}^{\bar{p}}{}_1)~,
\cr
\Omega_{p,}{}^p{}_1 &=& {1 \over 4} \big(\Omega_{p_1,p_2 p_3} \epsilon^{p_1 p_2 p_3} +
\Omega_{\bar{p}_1,\bar{p}_2 \bar{p}_3} \epsilon^{\bar{p}_1 \bar{p}_2 \bar{p}_3} - \Omega_{1,p}{}^p +\Omega_{\bar{1},p}{}^p~,
\cr
&+& \Omega_{+,+1}+\Omega_{+,+\bar{1}} +\Omega_{1,1\bar{1}}-\Omega_{\bar{1},1\bar{1}}\big)
-{1 \over 2}\big(\Omega_{p,}{}^p{}_{\bar{1}} - \Omega_{\bar{p},}{}^{\bar{p}}{}_1 \big)~,
\cr
\Omega_{-,-+} &=& -{1 \over 2}\big( \Omega_{1,-1}+\Omega_{\bar{1},-\bar{1}}+\Omega_{\bar{1},-1}+\Omega_{1,-\bar{1}} \big)~,
\cr
\Omega_{\bar{1},-\bar{1}} - \Omega_{1,-1} &=& \Omega_{-,p}{}^p-\Omega_{+,p}{}^p~,
\cr
\Omega_{-,1\bar{1}} &=&-\Omega_{-,p}{}^p~,
\cr
\Omega_{+,1\bar{1}} &=& \Omega_{+,p}{}^p~,
\cr
\Omega_{+,-+} &=&{1 \over 2}\big( \Omega_{1,-1}+\Omega_{\bar{1},-\bar{1}}+\Omega_{\bar{1},-1}+\Omega_{1,-\bar{1}} \big)~,
\cr
\Omega_{+,-1} &=&0~,
\cr
\Omega_{-,+1} &=&0~,
\eea
and
\bea
f^{-1} \partial_+ f &=& -{1 \over 2}\Omega_{-,-+}~,
\cr
f^{-1} \partial_- f &=& -{1 \over 2} \Omega_{-,-+}~,
\cr
f^{-1} \partial_1 f &=& -{1 \over 4}(\Omega_{+,+1}+\Omega_{-,-1})~,
\eea
and their complex conjugates.

The conditions on the geometry and fluxes that transform as  $(1,1)$ tensors under $SU(3)$ are
\bea
F_{+-1p \bar{q}} &=& \pm {1 \over 2} (\Omega_{p,+\bar{q}}+ \delta_{p \bar{q}} \Omega_{r,}{}^r{}_+)~,
\cr
F_{-1 \bar{1} p \bar{q}} &=& \mp {1 \over 4} (2 \Omega_{p,\bar{q}\bar{1}}+\Omega_{p,q_1 q_2} \epsilon^{q_1 q_2}{}_{\bar{q}})
\pm {1 \over 4} \delta_{p \bar{q}} (2 \Omega_{r,}{}^r{}_{\bar{1}}+\Omega_{r_1,r_2 r_3} \epsilon^{r_1 r_2 r_3})~,
\cr
F_{+1 \bar{1} p \bar{q}} &=& \pm {1 \over 4} (2 \Omega_{p,\bar{q} 1}-\Omega_{p,q_1 q_2} \epsilon^{q_1 q_2}{}_{\bar{q}})
\mp {1 \over 4} \delta_{p \bar{q}} (2 \Omega_{r,}{}^r{}_1-\Omega_{r_1,r_2 r_3} \epsilon^{r_1 r_2 r_3})~,
\cr
\Omega_{p,\bar{q}+} &=& \Omega_{\bar{q},p-}~,
\cr
2 \Omega_{p,\bar{q} \bar{1}} + \Omega_{p,q_1 q_2} \epsilon^{q_1 q_2}{}_{\bar{q}} &=& 2 \Omega_{\bar{q},p1} + \Omega_{\bar{q},\bar{q}_1
\bar{q}_2} \epsilon^{\bar{q}_1 \bar{q}_2}{}_p~,
\cr
2 \Omega_{p,\bar{q} 1} -  \Omega_{p,q_1 q_2} \epsilon^{q_1 q_2}{}_{\bar{q}} &=& 2 \Omega_{\bar{q},p \bar{1}} -\Omega_{\bar{q},\bar{q}_1
\bar{q}_2} \epsilon^{\bar{q}_1 \bar{q}_2}{}_p~,
\eea
and their complex conjugates.

The conditions on the geometry and fluxes that transform under the fundamental representation of $SU(3)$ are
\bea
F_{+-1 \bar{1}p} &=& \pm {1 \over 2} \big( \Omega_{p,+1}-\Omega_{p,-1}+\Omega_{+,1p}-{1 \over 2}
\Omega_{+,\bar{q}_1 \bar{q}_2} \epsilon^{\bar{q}_1 \bar{q}_2}{}_p \big)~,
\cr
F_{+-pq}{}^q &=& \pm {1 \over 2} \big(-\Omega_{p,-1}-\Omega_{p,+1}-\Omega_{+,1p}+{1 \over 2}
\Omega_{+,\bar{q}_1 \bar{q}_2} \epsilon^{\bar{q}_1 \bar{q}_2}{}_p \big)~,
\cr
F_{-1pq}{}^q &=&\pm \big( -{1 \over 2} \Omega_{\bar{1},\bar{q}_1 \bar{q}_2} \epsilon^{\bar{q}_1 \bar{q}_2}{}_p
+\Omega_{\bar{1},1p}-\Omega_{p,1\bar{1}}-\Omega_{p,q}{}^q \big)~,
\cr
F_{+\bar{1}pq}{}^q &=& \pm \big(-{1 \over 2} \Omega_{\bar{1},\bar{q}_1 \bar{q}_2} \epsilon^{\bar{q}_1 \bar{q}_2}{}_p
+\Omega_{p,-+}+\Omega_{\bar{1},1p}-\Omega_{p,q}{}^q\big)~,
\cr
F_{+-1\bar{q}_1\bar{q}_2} \epsilon^{\bar{q}_1 \bar{q}_2}{}_p &=& \mp \Omega_{p,+1}~,
\cr
F_{+-\bar{1}\bar{q}_1\bar{q}_2} \epsilon^{\bar{q}_1 \bar{q}_2}{}_p &=& \mp \Omega_{\bar{1},+p}~,
\cr
F_{-1 \bar{1} \bar{q}_1 \bar{q}_2} \epsilon^{\bar{q}_1 \bar{q}_2}{}_p &=& \pm \big(
-{1 \over 2} \Omega_{\bar{1},\bar{q}_1 \bar{q}_2} \epsilon^{\bar{q}_1 \bar{q}_2}{}_p
+\Omega_{\bar{1},1p} \big)~,
\cr
F_{+1 \bar{1} \bar{q}_1 \bar{q}_2} \epsilon^{\bar{q}_1 \bar{q}_2}{}_p &=& \pm \big(\Omega_{p,-+}-\Omega_{p,1\bar{1}}
+\Omega_{\bar{1},1p}-{1 \over 2} \Omega_{\bar{1},\bar{q}_1 \bar{q}_2} \epsilon^{\bar{q}_1 \bar{q}_2}{}_p \big)~,
\cr
\Omega_{p,-\bar{1}} &=& \Omega_{\bar{1},+p}~,
\cr
\Omega_{p,+\bar{1}} &=& -\Omega_{+,1p}+{1 \over 2}\Omega_{+,\bar{q}_1 \bar{q}_2}\epsilon^{\bar{q}_1 \bar{q}_2}{}_p
-\Omega_{p,+1}~,
\cr
\Omega_{1,-p} &=& \Omega_{p,+1}~,
\cr
\Omega_{1,+p} &=& \Omega_{p,-1}~,
\cr
\Omega_{1,\bar{1}p} &=& \Omega_{p,-+}-\Omega_{p,1\bar{1}}+\Omega_{\bar{1},1p}
-{1 \over 2}\Omega_{1,\bar{q}_1 \bar{q}_2} \epsilon^{\bar{q}_1 \bar{q}_2}{}_p
-{1 \over 2}\Omega_{\bar{1},\bar{q}_1 \bar{q}_2} \epsilon^{\bar{q}_1 \bar{q}_2}{}_p~,
\cr
\Omega_{1,1p} &=& -{1 \over 2}\Omega_{\bar{1},\bar{q}_1 \bar{q}_2}\epsilon^{\bar{q}_1 \bar{q}_2}{}_p
+\Omega_{\bar{1},1p} +{1 \over 2} \Omega_{1,\bar{q}_1 \bar{q}_2} \epsilon^{\bar{q}_1 \bar{q}_2}{}_p
-\Omega_{p,1\bar{1}}-\Omega_{p,q}{}^q~,
\cr
\Omega_{\bar{q},}{}^{\bar{q}}{}_p &=& \Omega_{\bar{q}_1,\bar{q}_2 \bar{1}} \epsilon^{\bar{q}_1 \bar{q}_2}{}_p~,
\cr
\Omega_{\bar{q}_1, \bar{q}_2 -} \epsilon^{\bar{q}_1 \bar{q}_2}{}_p &=& -\Omega_{p,+1}
+{1 \over 2}\Omega_{+,\bar{q}_1 \bar{q}_2} \epsilon^{\bar{q}_1 \bar{q}_2}{}_p
+\Omega_{p,-1}-\Omega_{+,1p}~,
\cr
\Omega_{\bar{q}_1,\bar{q}_2 +} \epsilon^{\bar{q}_1 \bar{q}_2}{}_p &=& \Omega_{p,+1}
-{1 \over 2}\Omega_{+,\bar{q}_1 \bar{q}_2} \epsilon^{\bar{q}_1 \bar{q}_2}{}_p
-\Omega_{p,-1}+\Omega_{+,1p}~,
\cr
\Omega_{\bar{q}_1, \bar{q}_2 1} \epsilon^{\bar{q}_1 \bar{q}_2}{}_p &=& -\Omega_{\bar{q}_1, \bar{q}_2 \bar{1}}
\epsilon^{\bar{q}_1 \bar{q}_2}{}_p~,
\cr
\Omega_{\bar{1},-p} &=& - \Omega_{+,1p}+{1 \over 2}\Omega_{+,\bar{q}_1 \bar{q}_2}\epsilon^{\bar{q}_1 \bar{q}_2}{}_p
-\Omega_{p,+1}~,
\cr
\Omega_{\bar{1},\bar{1}p} &=&\Omega_{p,-+}+\Omega_{\bar{1},1p}-\Omega_{\bar{1},\bar{q}_1 \bar{q}_2}
\epsilon^{\bar{q}_1 \bar{q}_2}{}_p - \Omega_{p,q}{}^q~,
\cr
\Omega_{-,1p} &=& {1 \over 2}\Omega_{-,\bar{q}_1 \bar{q}_2}\epsilon^{\bar{q}_1 \bar{q}_2}{}_p~,
\cr
\Omega_{-,-p} &=&  -\Omega_{\bar{1},\bar{q}_1 \bar{q}_2}\epsilon^{\bar{q}_1 \bar{q}_2}{}_p
-\Omega_{p,1\bar{1}}+2 \Omega_{\bar{1},1p} - \Omega_{p,q}{}^q~,
\cr
\Omega_{-,+p}&=&0~,
\cr
\Omega_{-,\bar{1}p} &=& - \Omega_{\bar{1},+p}-\Omega_{p,-1}-{1 \over 2}\Omega_{-,\bar{q}_1 \bar{q}_2}
\epsilon^{\bar{q}_1 \bar{q}_2}{}_p~,
\cr
\Omega_{+,-p}&=&0~,
\cr
\Omega_{+,+p}&=&2 \Omega_{\bar{1},1p}+2\Omega_{p,-+}-\Omega_{p,1\bar{1}}-\Omega_{\bar{1},\bar{q}_1
\bar{q}_2} \epsilon^{\bar{q}_1 \bar{q}_2}{}_p -\Omega_{p,q}{}^q~,
\cr
\Omega_{+,\bar{1}p} &=& -{1 \over 2}\Omega_{+,\bar{q}_1 \bar{q}_2} \epsilon^{\bar{q}_1 \bar{q}_2}{}_p~,
\eea
together with
\be
f^{-1} \partial_p f = {1 \over 2} \big(-\Omega_{p,-+}
+\Omega_{p,1\bar{1}}-2\Omega_{\bar{1},1p}+ \Omega_{\bar{1},\bar{q}_1 \bar{q}_2} \epsilon^{\bar{q}_1 \bar{q}_2}{}_p
+\Omega_{p,q}{}^q \big)~,
\ee
and their complex conjugates.

Lastly, the conditions on the geometry and fluxes that transform as  $(2,0)$ tensors under $SU(3)$ are
\bea
F_{+- \bar{q}_1 \bar{q}_2 (p} \epsilon^{\bar{q}_1 \bar{q}_2}{}_{q)} &=& \mp \Omega_{(p,|-|q)}~,
\cr
F_{-1 \bar{q}_1 \bar{q}_2 (p} \epsilon^{\bar{q}_1 \bar{q}_2}{}_{q)} &=& \pm \big(
\Omega_{(p,|1|q)} -{1 \over 2} \Omega_{(p,|\bar{q}_1 \bar{q}_2|}\epsilon^{\bar{q}_1 \bar{q}_2}{}_{q)} \big)~,
\cr
F_{+ \bar{1} \bar{q}_1 \bar{q}_2 (p} \epsilon^{\bar{q}_1 \bar{q}_2}{}_{q)} &=& \mp \big(
\Omega_{(p,|\bar{1}|q)} +{1 \over 2} \Omega_{(p,|\bar{q}_1 \bar{q}_2|}\epsilon^{\bar{q}_1 \bar{q}_2}{}_{q)} \big)~,
\cr
\Omega_{(p,|-|q)} &=& \Omega_{(p,|+|q)}~,
\eea
and their complex conjugates. As we have seen the above conditions considerably simplify when they are
written in terms of the fundamental forms of $G_2$.

\appendix{Null Structures}

\setcounter{subsection}{0}

\subsection{Null vectors and $SO(n)\ltimes \bR^n$ structures}
The stability subgroup in the special Lorentz group $SO(n+1,1)$ of a nowhere vanishing null vector $X$ is
$SO(n)\ltimes \bR^n$. Therefore, geometrically the structure of a Lorentzian manifold that admits
a non-vanishing null vector reduces to $SO(n)\ltimes \bR^n$. Topologically, the structure reduces further
to the maximal compact subgroup $SO(n)$.   Let $X$ be a non-vanishing null vector field on the spacetime.
It is always possible
to introduce a frame $e^+, e^-, e^i$ such that
\bea
ds^2= 2 e^+ e^-+\delta_{ij} e^i e^j~,~~~~X= e_+~,
\eea
where $e_A$ is the co-frame, $e^A(e_B)=\delta^A{}_B$. It is convenient to use the Lorentzian metric
to construct the associated null one-form $\kappa=e^-$ to $X$. Next consider the covariant derivative of $\kappa$, $\nabla \kappa$
with respect to the Levi-Civita connection of the Lorentzian metric. One way to determine the $SO(n)\ltimes \bR^n$ structures
is to decompose $\nabla \kappa$ under the irreducible representations of either the geometric structure group  $SO(n)\ltimes \bR^n$
or the topological structure group $SO(n)$. Since the representation of $SO(n)\ltimes \bR^n$ on the space of one-forms
is reducible but indecomposable,
a more refined characterization of the geometry can be achieved by decomposing $\nabla\kappa$ under the topological
structure group $SO(n)$. In particular, one finds that
\bea
\nabla \kappa= Y_1+Y_2+Y_3+Y_4+Y_5+Z_1+Z_2+Z_3~,
\eea
where
\bea
&&Y_1=\nabla_+\kappa_-~,~~~Y_2=\nabla_-\kappa_-~,~~~(Y_3)_i=\nabla_i\kappa_-~,~~~(Y_4)_i=\nabla_-\kappa_i~,~~~(Y_5)_i=\nabla_+\kappa_i
\cr
&&(Z_1)_{ij}=2\nabla_{[i}\kappa_{j]}~,~~~Z_2=\nabla^i\kappa_i~,~~~(Z_3)_{ij}=\nabla_{(i}\kappa_{j)}-{1\over n} \delta_{ij} \nabla^l\kappa_l~.
\eea
Clearly, if $\kappa$ is parallel, then all the classes vanish. If $X$ is Killing, then $Y_1=Y_2=Y_3+Y_4=Y_5=Z_2=Z_3=0$, and
similarly if $X$ is self-parallel, $Y_1=Y_5=0$. In all, there are $2^8$ possible structures.

\subsection{Null $U(n)\ltimes\bC^n$, Cauchy-Riemann and $SU(n)\ltimes\bC^n$ structures }\la{sun}

A $2n+2$-dimensional Lorentzian manifold with a $U(n)\ltimes\bC^n$-structure admits
a nowhere vanishing null one-form $\kappa$ and a three-form $\sigma=\kappa\wedge \omega$, where $\omega$ is a Hermitian form.
In an adapted basis, one has
\bea
\kappa=e^-~,~~~~\sigma=-ie^-\wedge\delta_{\a\bar\b} e^\a\wedge e^{\bar\b}~,~~~ds^2=2e^- e^++2 \delta_{\a\bar\b} e^\a e^{\bar\b}~.
\eea
It is worth pointing out that the stability subgroup of $e^-$ and $\omega$ in the special Lorentz group
$SO(2n+1,1)$ is not $U(n)\ltimes\bC^n$ and so the introduction of the null three-form $\sigma$ is necessary.

To determine the different $U(n)\ltimes\bC^n$ structures on the spacetime, we decompose $\nabla\kappa$ and $\nabla\sigma$
under representations of $U(n)$,  the maximal compact subgroup of $U(n)\ltimes\bC^n$, which is the topological structure
group of the spacetime\footnote{One may also consider decompositions under the $U(n)\ltimes\bC^n$ group. However,
the decomposition under $U(n)$ is more convenient since the irreducible representations are easy to identify.}.
It turns out that all independent structures can be found by considering the covariant derivative
of $\kappa$, $\nabla_A\kappa_B$, and the component,
$\nabla_A\omega_{ij}=\nabla_A \sigma_{-ij}-\nabla_A\kappa_-\, \omega_{ij}$, of $\nabla\sigma$. In particular
the $SO(2n)\ltimes\bR^{2n}$ classes which we have previously investigated are further decomposed as
\bea
&&Y_3=Y^{1,0}_3+ Y^{0,1}_3~,~~~Y_4=Y^{1,0}_4+ Y^{0,1}_4~,~~~Y_5=Y^{1,0}_5+Y^{0,1}_5~,~~~
\cr
&&Z_1=Z_1^{2,0}+Z_1^{0,2}+ \hat Z_1^{1,1}
-{i\over n}\omega\wedge ({\cal Z}_2-\bar {\cal Z}_2)~,~~~Z_2={\cal Z}_2+\bar {\cal Z}_2~,~~~{\cal Z}_2=\nabla^\a\kappa_\a~,
\cr
&& Z_3=Z_3^{2,0}+Z_3^{0,2}+\hat Z_3^{1,1}~.
\eea
In addition, there are classes associated with $\nabla\sigma$. The independent ones are
\bea
V_1^{2,0}=\nabla_+\omega_{\a\b}~,~~~~V_2^{2,0}=\nabla_-\omega_{\a\b}~.
\eea
and $W_1, W_2, W_3, W_4$ which are  associated with $\nabla_i\omega_{jk}$. The latter are related to those of Gray-Hervella
for almost Hermitian  manifolds, see \cite{gray}.

A Cauchy-Riemann structure is  a null $U(n)\ltimes \bC^n$ structure. A
Cauchy-Riemann structure determines an integrable distribution spanned by $e^-, e^{\bar\a}$. Using the torsion free
conditions, we observe that this requires that
\bea
\Omega_{[\a,\b]+}=\Omega_{+,+\a}=\Omega_{[\a,\b]\g}=\Omega_{[+,\b]\g}=0~.
\eea
In turn these give
\bea
\Omega_{\a,\b\g}=\Omega_{\a,\b+}=\Omega_{+,+\a}=\Omega_{+,\a\b}=0~.
\eea
Observe that in this setting $\kappa$ is not self-parallel. This in addition will require that $\Omega_{+,+-}=0$.
Therefore a Cauchy-Riemann structure is specified by the vanishing of the classes
\bea
W_1=W_2=0~,~~~V_1^{2,0}=V_2^{0,2}=0~,~~~Y_5=Z_1^{2,0}=Z_1^{0,2}=Z_3^{2,0}=Z_3^{0,2}=0~.
\eea
If $\kappa$ is self-parallel, then in addition $Y_1=0$.

The null $SU(n)\ltimes\bC^n$ structures can be investigated in a similar way. The associated
nowhere vanishing forms in the basis introduced for the $U(n)\ltimes\bC^n$ case are
\bea
\kappa=e^-~,~~~\sigma=e^-\wedge \omega~,~~~\rho=e^-\wedge \chi~,
\eea
where $\chi$ is the $SU(n)$-invariant holomorphic (n,0)-form. It turns out that the decomposition
of $\nabla\kappa$ is as in the $U(n)\ltimes\bC^n$ case above. In addition, one can also define
the classes $V_1$ and $V_2$. There are two new additional classes
\bea
V_3^{n-1,1}=\nabla_+({\rm Re}\,\chi)_{\a_1\dots\a_{n-1}\bar\b}~,~~~
V_4^{n-1,1}=\nabla_-({\rm Re}\,\chi)_{\a_1\dots\a_{n-1}\bar\b}~,
\eea
The remaining classes are determined by $\nabla_i\omega_{jk}$ and $\nabla_i\chi_{j_1\dots j_n}$. These
can be expressed in terms of the $W_1, W_2, W_3, W_4$ and $W_5$ classes of the $SU(n)$ structures, see e.g.~\cite{salamonb}.
The normalization which we use for these classes is that of appendix C  in \cite{commonsec}. In particular for the $SU(4)$ case which is relevant
to the results of this paper, we have
that
\bea
\tilde d\, \omega=W_1+W_3+{1\over 3} W_4\wedge\omega ~,
\cr
\tilde d\,{\rm Re}\chi=W_5\wedge {\rm Re} \chi+(-{1\over 3} W_1+{1\over2} W_2)\bar\wedge {\rm Im}\chi~,
\cr
\tilde d\,{\rm Im}\chi=W_5\wedge {\rm Im} \chi-(-{1\over 3} W_1+{1\over2} W_2)\bar\wedge {\rm Re}\chi~.
\la{exexsu4}
\eea
where $\tilde d$ denotes the exterior derivative evaluated along the transverse directions, $W_4$ is the Lee form of $\omega$, $W_5$
is the Lee form of ${\rm Re}\chi$.
It is clear that if $W_1=W_2=0$, then one has that
\bea
\tilde d\, \omega^3&=& W_4\wedge \omega^3~,
\cr
\tilde d\,\chi&=&W_5\wedge  \chi~.
\la{exexsu4b}
\eea
Observe that any Hermitian eight-dimensional manifold with an $SU(4)$ structure satisfies these conditions. As we have
seen supersymmetry imposes in addition that $W_4=W_5$. We refer to these  as {\it relatively balanced Hermitian $SU(4)$ structures}
because the difference of the two Lee forms vanishes. One consequence of (\ref{exexsu4b}) is that if $B$ admits
 such relatively balanced Hermitian structure, then $\tilde d W_4=\tilde d W_5$, is a trace-less (1,1)-form.

\subsection{Null $Spin(7)\ltimes\bR^8$ structures}

The null $Spin(7)\ltimes\bR^8$ structure is associated with the forms
\bea
\kappa=e^-~,~~~\sigma=e^-\wedge \phi~,~~~
\eea
where $\phi$ is the self-dual $Spin(7)$-invariant four-form.

The different $Spin(7)\ltimes\bR^8$ null structures can be determined by decomposing $\nabla\kappa$ and
$\nabla\sigma$ in $Spin(7)$ representations. It turns out that
\bea
\nabla\kappa=Y_1+\dots+Y_5+Z_1+\dots+Z_4~,
\eea
where
\bea
&&Y_1=\nabla_+\kappa_-~,~~~Y_2=\nabla_-\kappa_-~,~~~
(Y_3)_i=\nabla_i\kappa_-~,~~~(Y_4)_i=\nabla_-\kappa_i~,~~~(Y_5)_i=\nabla_+\kappa_i
\cr
&&(Z_1)_{ij}=d\kappa_{ij}|_{\bf 7}~,~~~(Z_2)_{ij}=d\kappa_{ij}|_{\bf 21}~,~~~Z_3=\nabla^i\kappa_i~,~~~
\cr
&&(Z_4)_{ij}=\nabla_{(i}\kappa_{j)}-{1\over 8} \delta_{ij} \nabla^l\kappa_l~,
\eea
where we have used the decomposition $\Lambda^2(\bR^8)=\Lambda_{\bf 7}\oplus \Lambda_{\bf 21}$ under
$Spin(7)$. It remains to investigate $\nabla\sigma$. It turns out that the remaining independent
structures are given by $\nabla_+\phi$, $\nabla_-\phi$ and $\nabla_i\phi_{j_1\dots j_4}$.
So we define
\bea
V_1=\nabla_+\phi_{j_1\dots j_4}~,~~~V_2=\nabla_-\phi_{j_1\dots j_4}~.
\eea
It is easy to see that both $V_1$ and $V_2$ lie in the fundamental seven-dimensional representation of $Spin(7)$.
Furthermore, $\nabla_i\phi_{j_1\dots j_4}$ determines two classes $W_1, W_2$ which are precisely those
expected for  eight-dimensional manifolds
with a $Spin(7)$ structure.

\appendix {$G_2$-structures in ten-dimensions}
\la{g2}

The $G_2$-structure we consider  is characterized by the existence of three one-forms
$e^+, e^-, e^1$ and a  three-form $\varphi$.  The metric can be written as
\bea
ds^2= 2 e^+ e^-+ (e^1)^2+ \delta_{ij} e^i e^j~.
\eea
The three one-forms span a trivial bundle in the decomposition $T^*M=I^3\oplus {\cal T}^*$ we have mentioned in section two.
The form $\varphi$ is the fundamental   $G_2$ three-form in the transverse directions ${\cal T}^*$.
Throughout, we use the notation of \cite{ugplgp}.
Following the analysis in \cite{gray} and using a unified notation for the three one-forms  as $\kappa_r$, $r=+,-1$, the different $G_2$ structures
are determined by decomposing the covariant derivatives $\nabla \kappa_r$  and $\nabla \varphi$
in $G_2$ representations. It turns out that it suffices to consider the classes
\bea
\nabla_r(\kappa_s)_t&\longleftrightarrow&  T_{rst}
\cr
\nabla_i(\kappa_s)_t&\longleftrightarrow& (V_1)_{st}
\cr
\nabla_r (\kappa_s)_i&\longleftrightarrow& (V_2)_{rs}
\cr
\nabla_i(\kappa_r)_j={1\over2} (\tilde d \kappa_r)_{ij}+\nabla_{(i}(\kappa_r)_{j)}&\longleftrightarrow& Z_r+F_r+T'_r+S_r
\cr
\nabla_r\varphi_{ijk}&\longleftrightarrow& Y_r~,
\cr
\nabla_i\varphi_{jkl}&\longleftrightarrow& X_1+X_2+X_3+X_4~,
\eea
where $X_1, X_2, X_3$ and $X_4$ are the usual $G_2$ classes in seven-dimensions.
The representations $T, T'$ and $X_1$ are singlets, $V_1, V_2, Z, Y$ and $X_4$ are 7-dimensional, $F$ and $X_2$
are 14-dimensional, and $S$ and $X_3$ are 27-dimensional.

One can write
\bea
\tilde d \varphi&\longleftrightarrow &X_1+X_3+X_4
\cr
\tilde d\varphi\wedge \varphi &\longleftrightarrow& X_1
\cr
\tilde d\star\varphi& \longleftrightarrow& X_2+X_4
\cr
\star(\tilde d\star\varphi)\wedge \star\varphi&\longleftrightarrow &X_4~,
\la{cocal}
\eea
where $\tilde d$ is the restriction of the exterior derivative along the transverse directions.
In particular, one finds
\bea
\tilde d\varphi&=&{1\over7} X_1\star\varphi+{3\over4} \theta\wedge \varphi-{1\over2} s_\Gamma\bar\wedge \star\varphi~,
\cr
\tilde d \star \varphi &=& X_2 \wedge \varphi + \theta \wedge \star \varphi
\eea
where $X_2 = \Pi_{{\bf{14}}} \star {\tilde{d}} \star \varphi$ and
$X_4=\theta_{\varphi}=-{1\over3} \star(\star \tilde d\varphi\wedge \varphi)$ is the Lee form,
\bea
&&X_1={1\over 4!} (\tilde d\varphi)_{i_1\dots i_4} \star\varphi^{i_1\dots i_4}
\cr
&&(s_\Gamma)_{ij}={1\over3!} (\tilde d\varphi)_{k_1k_2k_3(i} \star\varphi_{j)}{}^{k_1k_2k_3}+ {1\over 42}
\delta_{ij} (\tilde d\varphi)_{k_1k_2k_3 k_4} \star\varphi^{k_1k_2k_3 k_4}~.
\eea
Moreover we choose $s_\Gamma$ to represent $X_3$. The classes $X_1,\dots, X_4$ are analogous to the Fernandez-Gray
classes of seven-dimensional manifolds with a $G_2$ structure \cite{fgray}.


\begin{thebibliography}{00}
\addcontentsline{toc}{section}{References} \frenchspacing \small
\addtolength{\itemsep}{-4pt}


\bibitem{schwarz}
J.~H.~Schwarz, ``Covariant field equations of chiral N=2 D = 10
supergravity,'' Nucl.\ Phys.\ B {\bf 226} (1983) 269.

\bi{d3brane}
  G.~T.~Horowitz and A.~Strominger,
  ``Black strings and p-branes,''
  Nucl.\ Phys.\  B {\bf 360} (1991) 197.

  M.~J.~Duff and J.~X.~Lu,
  ``The selfdual type IIB superthreebrane,''
  Phys.\ Lett.\  B {\bf 273} (1991) 409.


\bi{bfhp}
M.~Blau, J.~Figueroa-O'Farrill, C.~Hull and G.~Papadopoulos,
``A new maximally supersymmetric background of IIB superstring theory,''
JHEP {\bf 0201} (2002) 047
[arXiv:hep-th/0110242].

``Penrose limits and maximal supersymmetry,''
Class.\ Quant.\ Grav.\  {\bf 19} (2002) L87
[arXiv:hep-th/0201081].






\bibitem{maldacena}
  J.~M.~Maldacena,
  ``The large N limit of superconformal field theories and supergravity,''
  Adv.\ Theor.\ Math.\ Phys.\  {\bf 2} (1998) 231
  [Int.\ J.\ Theor.\ Phys.\  {\bf 38} (1999) 1113]
  [arXiv:hep-th/9711200].

  O.~Aharony, S.~S.~Gubser, J.~M.~Maldacena, H.~Ooguri and Y.~Oz,
``Large N field theories, string theory and gravity,''
Phys.\ Rept.\  {\bf 323}, 183 (2000)
[arXiv:hep-th/9905111].


\bibitem{llmsol}
H.~Lin, O.~Lunin and J.~M.~Maldacena,
``Bubbling AdS space and 1/2 BPS geometries,''
JHEP 0410(2004)025; arXiv:hep-th/0409174


\bi{hjg}
 J.~B.~Gutowski and H.~S.~Reall,
  ``Supersymmetric AdS(5) black holes,''
  JHEP {\bf 0402} (2004) 006
  [arXiv:hep-th/0401042].

  ``General supersymmetric AdS(5) black holes,''
  JHEP {\bf 0404} (2004) 048
  [arXiv:hep-th/0401129].

\bibitem{jgjg}
  J.~P.~Gauntlett, J.~B.~Gutowski and N.~V.~Suryanarayana,
  ``A deformation of $AdS_5 \times S^5$,''
  Class.\ Quant.\ Grav.\  {\bf 21} (2004) 5021
  [arXiv:hep-th/0406188].

\bibitem{ahn}
  C.~H.~Ahn and J.~F.~Vazquez-Poritz,
  ``Deformations of flows from type IIB supergravity,''
  Class.\ Quant.\ Grav.\  {\bf 23} (2006) 3619
  [arXiv:hep-th/0508075].



\bibitem{ugjggpa}
U.~Gran, J.~Gutowski and G.~Papadopoulos,
``The spinorial geometry of supersymmetric IIB backgrounds,''
Class.\ Quant.\ Grav.\  {\bf 22} (2005) 2453
[arXiv:hep-th/0501177].




\bibitem{ugjggpb}
U.~Gran, J.~Gutowski and G.~Papadopoulos,
``The $G_2$ spinorial geometry of supersymmetric IIB backgrounds,''
Class.\ Quant.\ Grav.\  {\bf 23} (2006) 143
[arXiv:hep-th/0505074].

\bibitem{sw}
  J.~H.~Schwarz and P.~C.~West,
  ``Symmetries and transformations of chiral N=2 D=10 supergravity,''
  Phys.\ Lett.\  B {\bf 126} (1983) 301.


\bibitem{howe}
  P.~S.~Howe and P.~C.~West,
  ``The complete N=2, D=10 supergravity,''
  Nucl.\ Phys.\  B {\bf 238} (1984) 181.


\bibitem{jguggp}
  J.~Gillard, U.~Gran and G.~Papadopoulos,
  ``The spinorial geometry of supersymmetric backgrounds,''
  Class.\ Quant.\ Grav.\  {\bf 22} (2005) 1033
  [arXiv:hep-th/0410155].




\bibitem{gutgaunfive} J. P. Gauntlett and J. B. Gutowski,
``All supersymmetric solutions of minimal gauged supergravity in
five-dimensions," Phys. Rev. {\bf{D68}} (2003) 105009;
hep-th/0304064.

\bibitem{chamemp} A. Chamblin, R. Emparan, C. V. Johnson and R. C.
Myers, ``Charged AdS black holes and catastrophic holography,"
Phys. Rev. {\bf{D60}} (1999) 064018; hep-th/9902170.


\bibitem{iibsyst}
  U.~Gran, J.~Gutowski, G.~Papadopoulos and D.~Roest,
  ``Systematics of IIB spinorial geometry,''
  Class.\ Quant.\ Grav.\  {\bf 23} (2006) 1617
  [arXiv:hep-th/0507087].


\bi{gray}
A. Gray and L.M. Hervella, ``The sixteen classes of
 almost Hermitian manifolds and their linear invariants'',
 Ann.\ Mat.\ Pura\ e \ Appl.\ {\bf 282} (1980) 1.


\bi{salamonb} S. Chiossi and S. Salamon, ``The intrinsic torsion of $SU(3)$
 and $G_2$ structures'', Diff. Geom., Valencia 2001, World Sci. 2002, 115
 [arXiv:math.DG/0202282].


\bi{swann}
R. Cleyton and A. Swann, ``Cohomogeneity-one $G_2$-structures'', [arXiv:math/0111056].



\bibitem{paptown}
  G.~Papadopoulos and P.~K.~Townsend,
  ``Intersecting M-branes,''
  Phys.\ Lett.\  B {\bf 380} (1996) 273
  [arXiv:hep-th/9603087].

\bibitem{tseytlinh}
  A.~A.~Tseytlin,
  ``Harmonic superpositions of M-branes,''
  Nucl.\ Phys.\  B {\bf 475} (1996) 149
  [arXiv:hep-th/9604035].









\bibitem{ugplgp}
U.~Gran, P.~Lohrmann and G.~Papadopoulos,
``The spinorial geometry of supersymmetric heterotic string backgrounds,''
JHEP {\bf 0602} (2006) 063
[arXiv:hep-th/0510176].







\bi{jfgpa}
 J.~Figueroa-O'Farrill and
G.~Papadopoulos,
``Maximally supersymmetric solutions of ten-dimensional
and eleven-dimensional supergravities,''
JHEP {\bf 0303} (2003) 048:
[arXiv:hep-th/0211089].

``Pluecker-type relations for orthogonal planes,''
[arXiv:math.ag/0211170].




\bi{nps} Th. Friedrich, I. Kath, A. Moroianu and U. Semmelmann, ``Nearly parallel $G_2$ structures,'' J. Geom. Phys.
{\bf 23} (1997), no 3-4, 259-286.












\bibitem{tsimpis}
G.~Papadopoulos and D.~Tsimpis, ``The holonomy of IIB
supercovariant
connection,'' Class.\ Quant.\ Grav.\  {\bf 20} (2003) L253
[arXiv:hep-th/0307127].




\bibitem{hull}
  C.~Hull,
  ``Holonomy and symmetry in M-theory,''
  arXiv:hep-th/0305039.

\bibitem{duff}
  M.~J.~Duff and J.~T.~Liu,
  ``Hidden spacetime symmetries and generalized holonomy in M-theory,''
  Nucl.\ Phys.\  B {\bf 674} (2003) 217
  [arXiv:hep-th/0303140].






\bibitem{chen}
  B.~Chen {\it et al.},
  ``Bubbling AdS and droplet descriptions of BPS geometries in IIB
  supergravity,''
  arXiv:0704.2233 [hep-th].

 A.~Donos,
  ``BPS states in type IIB SUGRA with SO(4) x SO(2)(gauged) symmetry,''
  arXiv:hep-th/0610259.


  \bibitem{preons}
  U.~Gran, J.~Gutowski, G.~Papadopoulos and D.~Roest,
  ``N = 31 is not IIB,''
  JHEP {\bf 0702} (2007) 044
  [arXiv:hep-th/0606049].



\bibitem{typeI}
  U.~Gran, G.~Papadopoulos, D.~Roest and P.~Sloane,
  ``Geometry of all supersymmetric type I backgrounds,''
  arXiv:hep-th/0703143.


  \bibitem{jfof}
  J.~Figueroa-O'Farrill, E.~Hackett-Jones and G.~Moutsopoulos,
  ``The Killing superalgebra of ten-dimensional supergravity backgrounds,''
  arXiv:hep-th/0703192.



\bibitem{commonsec}
  U.~Gran, P.~Lohrmann and G.~Papadopoulos,
  ``Geometry of type II common sector N = 2 backgrounds,''
  JHEP {\bf 0606} (2006) 049
  [arXiv:hep-th/0602250].




  \bi{fgray} M. Fernandez and A. Gray,
 ``Riemannian manifolds with structure group $G_2$,''
 Ann. Mat. Pura Appl (4) 32 (1982), 19-45. 1.


\bibitem{romans}
  L.~Castellani and L.~J.~Romans,
  ``N=3 and N=1 supersymmetry in a new class of solutions for D = 11
  supergravity,''
  Nucl.\ Phys.\  B {\bf 238} (1984) 683.

  \bibitem{romansw}
  L.~Castellani, L.~J.~Romans and N.~P.~Warner,
  ``A classification of compactifying solutions for D = 11 supergravity,''
  Nucl.\ Phys.\  B {\bf 241}, 429 (1984).



\bi{aloff} S.~Aloff and N.~R.~Wallach, ``An infinite family of distinct 7-manifolds admitting positively curved riemannian structures,'' Bull.
Am. Math. Soc. {\bf 81} (1975), 93-97.


\bibitem{duffpope}
  M.~J.~Duff, B.~E.~W.~Nilsson and C.~N.~Pope,
  ``Kaluza-Klein Supergravity,''
  Phys.\ Rept.\  {\bf 130} (1986) 1.

\bibitem{castellani}
  L.~Castellani, A.~Ceresole, R.~D'Auria, S.~Ferrara, P.~Fre and M.~Trigiante,
  ``G/H M-branes and $AdS_{p+2}$ geometries,''
  Nucl.\ Phys.\  B {\bf 527} (1998) 142
  [arXiv:hep-th/9803039].

  \bi{boyer}C.~P.~Boyer, K.~Galicki, B.~M.~Mann, and E.~Rees, ``Compact 3-Sasakian 7-manifolds with arbitrary second Betti
  number,'' Invent. Math. {\bf 131} (1998), 321-344.

  \bi{gsalamon} K.~Galicki and S.~Salamon, ``On Betti numbers of 3-Sasakian manifolds,'' Geom. Ded {\bf 63} (1996), 45-68.



\bi{ujg} U.~Gran, J.~Gutowski and G.~Papadopoulos, in preparation.
\end{thebibliography}
\end{document}